\documentclass[aip,cha,reprint,a4paper,nofootinbib]{revtex4-1}

\usepackage[margin=1in]{geometry} 
\usepackage{amsmath,amsthm,amssymb}
\usepackage[margin=1in]{geometry} 
\usepackage{amsmath,amsthm,amssymb}
\usepackage{booktabs}

\setcitestyle{numbers,square}

\usepackage[T1]{fontenc} 
\usepackage[utf8]{inputenc} 
\usepackage{lmodern} 
\usepackage{graphicx}
\usepackage{subfigure}
\usepackage{xcolor}
\graphicspath{ {New_images/} }
\usepackage{hyperref} 
\usepackage{soul}

\usepackage{lineno}

\begin{document}
\title{Citations or dollars? Early signals of a firm's research success}

\author{Shuqi Xu}
\affiliation{Institute of Fundamental and Frontier Sciences, University of Electronic Science and Technology of China, Chengdu 610054, PR China}
\affiliation{Yangtze Delta Region Institute (Huzhou), University of Electronic Science and Technology of China, Huzhou 313001, PR China}

\author{Manuel S. Mariani}
\email{manuel.mariani@business.uzh.ch}
\affiliation{Institute of Fundamental and Frontier Sciences, University of Electronic Science and Technology of China, Chengdu 610054, PR China}
\affiliation{URPP Social Networks, University of Zurich, CH-8050 Zurich, Switzerland}

\author{Linyuan Lü}
\email{linyuan.lv@uestc.edu.cn}
\affiliation{Institute of Fundamental and Frontier Sciences, University of Electronic Science and Technology of China, Chengdu 610054, PR China}
\affiliation{Yangtze Delta Region Institute (Huzhou), University of Electronic Science and Technology of China, Huzhou 313001, PR China}
\affiliation{Beijing Computational Science Research Center, Beijing 100193, PR China}

\author{Lorenzo Napolitano} 
\affiliation{Istituto dei Sistemi Complessi  (ISC-CNR), UoS  Sapienza, Rome I-00185, Italy}
\affiliation{European Commission, Joint Research Centre (JRC), Seville 41092, Spain}

\author{Emanuele Pugliese}
\affiliation{Istituto dei Sistemi Complessi  (ISC-CNR), UoS Sapienza, Rome I-00185, Italy}
\affiliation{European Commission, Joint Research Centre (JRC), Seville 41092, Spain}

\author{Andrea Zaccaria}
\affiliation{Istituto dei Sistemi Complessi  (ISC-CNR), UoS Sapienza, Rome I-00185, Italy}

\begin{abstract}
Scientific and technological progress is largely driven by firms in many domains, including artificial intelligence and vaccine development. However, we do not know yet whether the success of firms' research activities exhibits dynamic regularities and some degree of predictability.
By inspecting the research lifecycles of 7,440 firms, we find that the economic value of a firm's early patents is an accurate predictor of various dimensions of a firm's future research success. At the same time, a smaller set of future top-performers do not generate early patents of high economic value, but they are detectable via the technological value of their early patents. 
Importantly, the observed predictability cannot be explained by a cumulative advantage mechanism, and the observed heterogeneity of the firms' temporal success patterns markedly differs from patterns previously observed for individuals' research careers.
Our results uncover the dynamical regularities of the research success of firms, and they could inform managerial strategies as well as policies to promote entrepreneurship and accelerate human progress.
\end{abstract}

\maketitle
\onecolumngrid

\section{Introduction}
In most technological sectors, corporate actors are the main drivers of innovation. For example, in the artificial intelligence (AI) domain, recent years have witnessed a dramatic increase in corporate expenditures on related research projects~\cite{idc2019worldwide}. These efforts have led to outstanding breakthroughs, such as the detection of proteins' 3D structure by DeepMind's AlphaFold~\cite{tunyasuvunakool2021highly}, as well as many failed products, such as Google glasses. During the ongoing pandemic, more COVID-19 vaccines are being developed by companies than by academic actors~\cite{le2020evolution}. Because of the prominent role played by companies for scientific and technological progress, understanding the regularities and predictability of firms' research success is vital for diverse players. It can indeed help managers to identify effective innovation strategies~\cite{hauser2006research} as well as high-potential investment opportunities~\cite{nanda2020persistent}, and policymakers to design effective policies that promote entrepreneurship and accelerate human progress~\cite{guzman2015silicon}.

However, potential regularities behind the success of the research outputs produced by corporate actors and its predictability remain unknown.
Most recent efforts to understand the success dynamics of research actors have indeed focused on individual scientists and teams of scientists or inventors~\cite{sinatra2016quantifying,liu2018hot,wang2019early,yin2019quantifying,wang2021science}.
Like scientific discoveries are linked to the previous body of knowledge through citations between academic papers~\cite{fortunato2018science,wang2021science}, novel inventions are linked to previous ones via citations between the corresponding patents~\cite{jaffe2019patent}.
Despite differences between the nature of scientific and patent citations~\cite{jaffe2019patent}, this compelling analogy has recently unveiled common patterns behind the dynamics of scientific and technological innovation. Common patterns exist, for example, in how successful scientific and technological research build on prior knowledge~\cite{uzzi2013atypical,mukherjee2017nearly,shi2019science,pugliese2019coherent,pugliese2019unfolding}, how the impact of papers and patents evolves over time~\cite{wang2013quantifying,higham2017fame}, and how team size predicts research impact and disruptiveness~\cite{wu2019large}. 
These studies emphasize the similarities between scientific and technological innovation, and they point to the potential benefits of patent analysis to understand the dynamics of firms' research success.

\begin{figure}[t]
\centering
\includegraphics[scale=0.38]{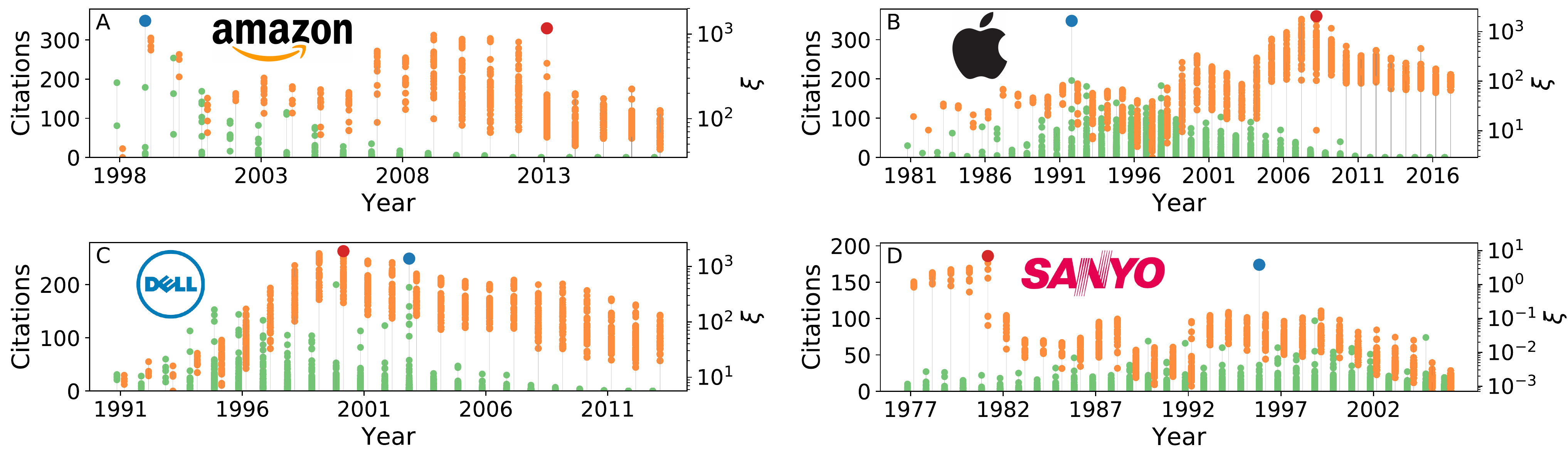}
\caption{\textbf{The research lifecycles of firms from their patents.} Inspired by recent studies on the dynamics of human achievements~\cite{sinatra2016quantifying,janosov2020success}, we view each firm as the temporal sequence of its issued patents, which we refer to as the firm's \textit{lifecycle}. Each patent is characterized by a metric of technological value (based on its number of received citations, denoted by green dots) and economic value ($\xi$, based on the firm's stock-price movements following the patent's announcement, represented by orange dots). 
The four panels show the patenting lifecycles of four major firms, \textit{Amazon}, \textit{Apple}, \textit{Dell}, and \textit{Sanyo}. This figure illustrates that the highest-cited patents (blue) and patents with the highest economic value (red) could emerge at different stages over firms' lifecycles.
}
\label{fig:firmCareer}
\end{figure}

Inspired by recent studies on scientists’ careers~\cite{sinatra2016quantifying,liu2018hot}, we represent a firm’s research lifecycle as the time-ordered sequence of its issued patents (see Fig.~\ref{fig:firmCareer}). Based on this representation, we ask the previously-unexplored questions: Do firms exhibit similar research success patterns as academic actors? Is the firms' future research success predictable from their earliest outputs? Which mechanisms lie behind the observed predictability?
An obstacle toward answering these questions is the ambivalence of the success of patents from the applicant firm's standpoint.
Quantitative studies of science often define the scientific success of a paper as a one-dimensional construct determined by its received citations~\cite{wang2013quantifying,fortunato2018science,wang2021science}.
However, defining the success of a patent as its received citations would only capture its technological value, but not its economical value which drives firms investment decisions~\cite{kogan2017technological,stoffman2020small}.

To overcome this obstacle, we analyse the patenting history of $7,440$ firms in the United States Patent and Trademark Office (USPTO) dataset from 1926 to 2017~\cite{Patent_CRSP}.
A recently-collected dataset~\cite{kogan2017technological} offers the unique opportunity to quantify simultaneously both the technological and the economic value of firms' patents via metrics based on the number of citations received~\cite{hall2005market} and the firms' stock-price movements following the patent's announcement~\cite{kogan2017technological}. To compare patents issued in different years, we normalize both metrics by requiring that the score of a patent is not biased by its issuing year~\cite{waltman2016review,mariani2019early}. We aim to quantify the predictability of firms' research success, and understand the different implications of patents' economic and technological value for a firm's research success.

We find that the economic value of a firm's early patents is predictive of the economic and technological value of its later patents.
On the other hand, the technological value of a firm's early patents is only predictive of the technological value of the firm's subsequent patents, but not of their future economic value.
To test potential mechanisms behind these findings, we perform a matched pair analysis~\cite{li2019early,alshebli2018preeminence,wang2019early}. The results provide evidence in favor of a fitness explanation where early success is a manifestation of the capability to produce high-value patents, and against a pure competitive advantage explanation where future success is caused by early success alone. 
Among firms without top-economic value patents in the early stage, ``hidden gem'' firms that are later granted high-economic value patents differ from those that are not (i.e., ``non-top'' firms) in the technological value of their early patents.
The typical lifecycles of hidden gem firms markedly differs from that of ``predictable'' firms that are among the top ones by economic value in both the early and late stage. This further reveals the non-random timing of a firm's best research markedly differs from the random timing of scientists' highest-impact papers~\cite{sinatra2016quantifying}, which calls for new models to describe the dynamics of firms' research success.

\section{Results}
\subsection{Quantifying the value of patents and firms \label{sec:Result_quantify}}

\begin{figure}[t]
\centering
\includegraphics[scale=0.4]{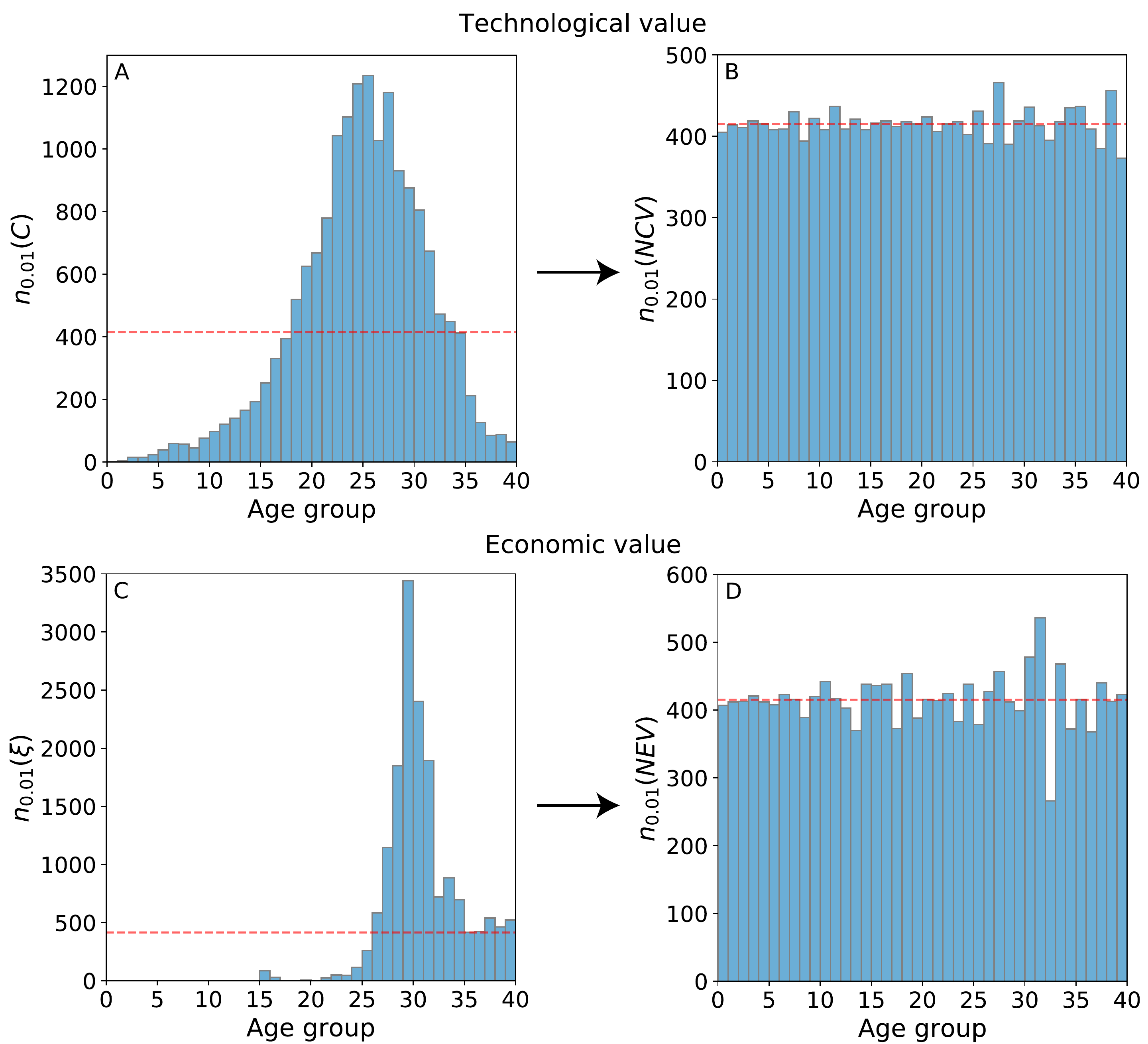}
\caption{\textbf{Normalizing patent-level technological and economic value metrics.} Age distribution of the top-1\% patents ranked by the citation count (panel A), economic value $\xi$ (panel C), and the proposed normalized value metrics (panels B and D). We divide all patents into 40 equally-sized groups by age and show the number of top-1\% patents from each age group~\cite{mariani2019early}. The dashed red line denotes the expected value for an age-unbiased ranking, $0.01\,N/40$, where $N$ denotes the total number of patents of all firms. The distributions for the raw value metrics ($C$ and $\xi$) are strongly biased by patent age (panels A and C), which makes them unsuitable to study the temporal patterns of firms. By contrast, the two normalized value metrics (\emph{NCV} and \emph{NEV}) exhibit a flat temporal profile (panels B and D).}
\label{fig:timebias}
\end{figure}

We start by defining the metrics of economic and technological value at the patent and firm level. We consider two dimensions of patents' value: their \textit{technological value} and their \textit{economic value}.
To quantify the technological value of a patent, we measure its number of received citations~\cite{hall2005market}. 
A potential shortcoming of the citation count (even when restricted to a $10$-year temporal window~\cite{sinatra2016quantifying}) is its strong bias by patent age (see Fig.~\ref{fig:timebias}A and Fig.~S1 in SI). 
To eliminate this bias, we compare each patent's citation count only against the citation counts of patents issued in the same year. Hence, we define the \textit{age-normalized citation value} (\emph{NCV}) of a patent as its ranking position by citation count among all patents issued in the same year (see Fig.~\ref{fig:timebias}B and Methods for details).

To quantify the economic value of a patent, we rely on a recent measure based on the firm stock price movement over a narrow time window after the patent is issued~\cite{stoffman2020small}. The core idea of this metric (denoted as $\xi$) is that the market's reaction to a patent is a combination of the dollar value of the patent and the investors' ex-ante probability assessment on the patent's success~\cite{stoffman2020small} (see Methods for details).
Differently from the time-varying citations of patents, a patent's $\xi$ is determined shortly after the patent's issuance and does not change over time. Similarly to citation count, the economic value metric $\xi$ also exhibits strong bias by patent age, as shown in Fig.~\ref{fig:timebias}C.
Again, to prevent this bias from influencing our firm-level results, we define the \textit{age-normalized economic value} (\emph{NEV}) of a patent as its ranking position by $\xi$ among all patents issued in the same year (see Fig.~\ref{fig:timebias}D and Methods for details).

Technological and economic value do not always coincide. A patent may represent a major technical advance, but its announcement might fail to restrict competition or attract the attention of investors, thereby generating a modest impact on the company's stock price.
For example, patent US$3728480$ from \textit{Sanders Associates} (see Table \ref{tab:significant}) reported the invention of the first video game that could be played on a home television. This can be considered as a substantial technological advance compared to computer games, and the patent was highly cited, resulting in a high technological value ($NCV=0.98$). At the same time, the patent failed to capture market interest shortly after its issuance, likely because of the recession in the cable TV industry at that time\footnote{\url{http://www.pong-story.com/sanders.htm}}, which resulted in a low economic value ($NEV=0.17$). 
We refer to Tables~\ref{tab:significant} and S1 in SI for the $NCV$ and $NEV$ of a set of expert-selected historically significant patents~\cite{strumsky2015identifying}, and to Tables S2--S3 for a list of top patents by \emph{NCV} and \emph{NEV}, respectively.

Overall, the Pearson correlation between patents' technological and economic value is as low as $r(NCV,NEV)=0.09$, and the correlation between the two non-normalized variables is also low ($r(c,\xi)=0.09$, see Fig.~S2 in SI). To explain the discrepancy of our finding and previous claims of high positive correlation between technological and economic value~\cite{cremers1999citation, hall2005market,kogan2017technological}, we show that such correlation increases as patents are grouped into increasingly-large sets of patents with a similar citation value (see Fig.~S2 in SI). Therefore, whereas previous works demonstrated that groups of patents with higher citation impact exhibit higher economic value~\cite{kogan2017technological}, the low correlation reported here indicates that there is little predictability of economic value from citation value at the individual patent level.

To quantify the research success of a given firm, one could average or sum the value of all its patents. However, it is well-known that patents' quality is highly heterogeneous~\cite{silverberg2007size}, and prior works placed a greater emphasis on a firm's most prominent innovation than on ordinary innovations~\cite{ahuja2001entrepreneurship,fleming2003navigating,dunlap2010story}. For this reason, we focus on a firm's patents with the highest technological and economic value, which we refer to as its technological and economic hit~\cite{ahuja2001entrepreneurship,srivastava2011relational}, respectively. 
The two hits coincide for a minority of firms, which account for $2\%$ of the analyzed firms (see Fig.~S3 in SI for the correlation details), and we show below that the value of early economic and technological hits have substantially different implications for the firms' future research success.

Based on the hits, we define the two dimensions of the innovation value of a given firm $\alpha$: its technological value ($TV$) and economic value ($EV$). We define the $TV$ and $EV$ of a given firm as the technological value of the firm's technological hit and the economic value of its economic hit, respectively. In formulas, $TV_{\alpha}=max_{i\in P_{\alpha}}\{NCV_i\}$ and $EV_{\alpha}=max_{i\in P_{\alpha}}\{NEV_i\}$, where $P_{\alpha}$ denotes the set of patents that were granted to firm $\alpha$. Note that to simplify exposition, in the following, we refer to a ``firm's value'' as a shorthand for its innovation value, i.e. the value of its patents. This should not be confused with the firm's stock price or other measures of firm's performance, which are not considered here.

We divide firms into three groups according to their technological value and economic value. Specifically, we consider the top-$5\%$ firms as high-value firms, the bottom-$35\%$ as low-value  firms, and the intermediate $60\%$ as medium-value firms. 
All our results do not strongly depend on the exact choice of these separation thresholds (see Figs.~S14 and S15 in SI).
These three groups of firms exhibit markedly different productivity (in terms of number of issued patents) and value dynamics (see SI Fig.~S4). High-value firms exhibit a sustained advantage over medium and low-value firms in terms of both productivity and value. This gap is evident even in the very early stage. Motivated by this finding, in the following, we will test whether early activity data can be used to predict firms' future value.

\begin{table}[t]
\small
  \centering
  \caption{The different technological value (\emph{NCV}) and economic value (\emph{NEV}) of five historically significant patents. Among the historically significant patents identified by Strumsky and Lobo~\cite{strumsky2015identifying}, we show a sample of five patents whose \emph{NCV} differ from the \emph{NEV}. We refer to Table~S1 in SI for the value metrics of all 31 significant patents.}
    \begin{tabular}{|r|l|l|r|r|p{4cm}|}
    \hline
    \multicolumn{1}{|l|}{Patent \#} & Issue year & Applicant firm  & \multicolumn{1}{l|}{\emph{NCV}} & \multicolumn{1}{l|}{\emph{NEV}} & Title/description \\
    \hline
    2895584 & 1959  & INTERNATIONAL BUSINESS MACHS COR  & 0.77  & 0.96  & Selectric typewriter printing head \\
    3728480 & 1973  & SANDERS ASSOCIATES INC & 0.98  & 0.17  & First video game \\
    3821715 & 1974  & GENERAL ELECTRIC CO & 1.00  & 0.83  & Intel 4004 microprocessor \\
    4504982 & 1985  & OPTICAL RADIATION CORP  & 0.99  & 0.41  & An intraocular lens for permanent implantation into a human eye \\
    6469012 & 2002  & PFIZER INC  & 0.66  & 0.99  & Viagra \\
    \hline
    \end{tabular}%
  \label{tab:significant}%
\end{table}%

\subsection{Early economic value predicts future research success}

We start by examining whether firms' early value predicts future research success.
To this end, we split each firm's research lifecycle into a $5$-year \textit{initial window} of early activity and a \textit{later window} composed of all its remaining years. Our results are qualitatively unchanged for different choices of the initial period's duration and the later period's duration (see SI S3.4). A firm's \textit{early technological} (\textit{economic}) \textit{value} is defined as the technological (economic) value of the early technological (economic) hit (i.e., the highest-value patent among the patents issued within the initial $5$-year window). We can define firms' subsequent technological and economic value in a similar way.

We find a strong predictability: firms among the top-$5\%$ by early economic value are $21.9$ times more likely to be among the top-$5\%$ by subsequent economic value than the other firms; firms among the top-$5\%$ by early technological value are $5.1$ times more likely to be among the top-$5\%$ by subsequent technological value than the other firms (see Fig.~S5 in SI). 
These initial findings motivate the question: Is early technological or economic value more predictive of firms' subsequent research success? We answer this question by first quantifying the predictive power of both variables, and then mimicking an experiment by creating treatment and control groups of firms that differ by their early technological or economic value.

To quantify the predictive power of firms' early technological and economic value, we study a set of classification problems where we use information on firms' early patents to predict which firms will subsequently be among the top-$5\%$ by two dimensions of future success: the technological value of the future technological hit (i.e., the highest-value patent among the patents issued in the late window), and the economic value of the future economic hit. 
Based on the literature, we consider various metrics of firms' early performance that might be predictive of future success: not only the firms' early technological and economic value, but also their early productivity (in terms of the total number of early patents)~\cite{ahuja2001technological,zhang2020foot}, total citations of early patents~\cite{trajtenberg1990penny,turkina2019regional}, and other aggregate measures of early patent value (in terms of cumulative $\xi$, \emph{NCV}, and \emph{NEV}).
For each of these early performance metrics, we measure various predictive accuracy metrics, including precision, recall, area under the precision-recall curve~\cite{powers2011evaluation}, for a Na\"ive Bayes Classifier that classifies a firm as successful if and only if it is among the top-$z\%$ by the metric, where $z$ is a parameter that can be tuned to achieve a desired value of recall (see SI S3.2).

\begin{figure}[t]
\centering
\includegraphics[scale=0.37]{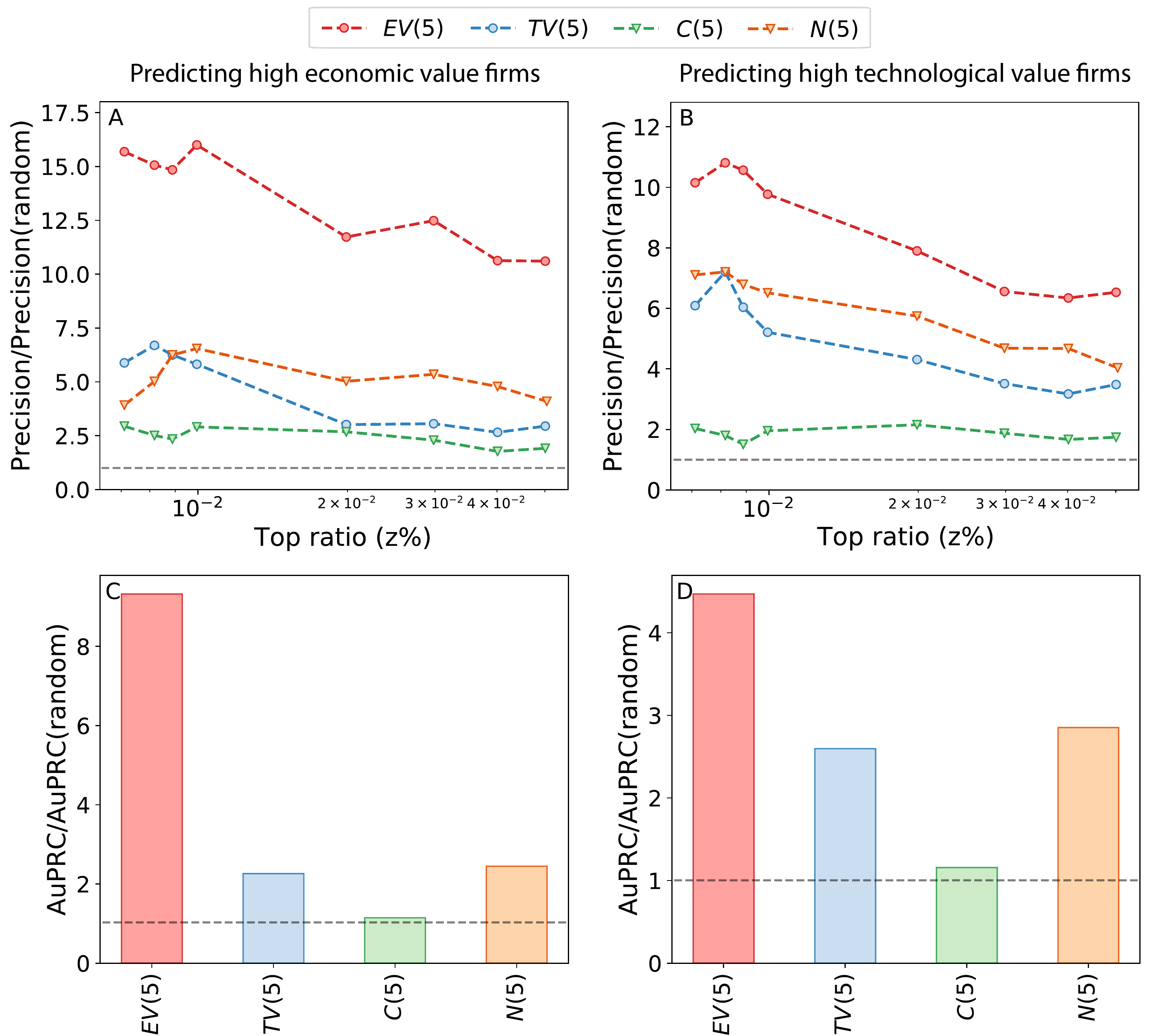}
\caption{\textbf{Predicting top-firms from their early patents.} We evaluate the predictive power of classifiers based on the top-$z$\% firms by various metrics of their early patents (within the earliest five years after their first patent issuance): economic value ($EV$), technological value ($TV$), total number of citations ($C$) and total number of patents ($N$).
The $EV$-based classifier outperforms the others in predicting top-$z\%$ firms by late economic value (A, C), and late technological value (B, D). Panels A, B refer to the classifiers' precision normalized by the precision of a random guess as a function of $z\%$; panels C, D refer to the area under the precision-recall curve (\emph{AuPRC}) 
}
\label{fig:Predict-final-removeEarly}
\end{figure}

\begin{figure}[t]
\centering
\includegraphics[scale=0.37]{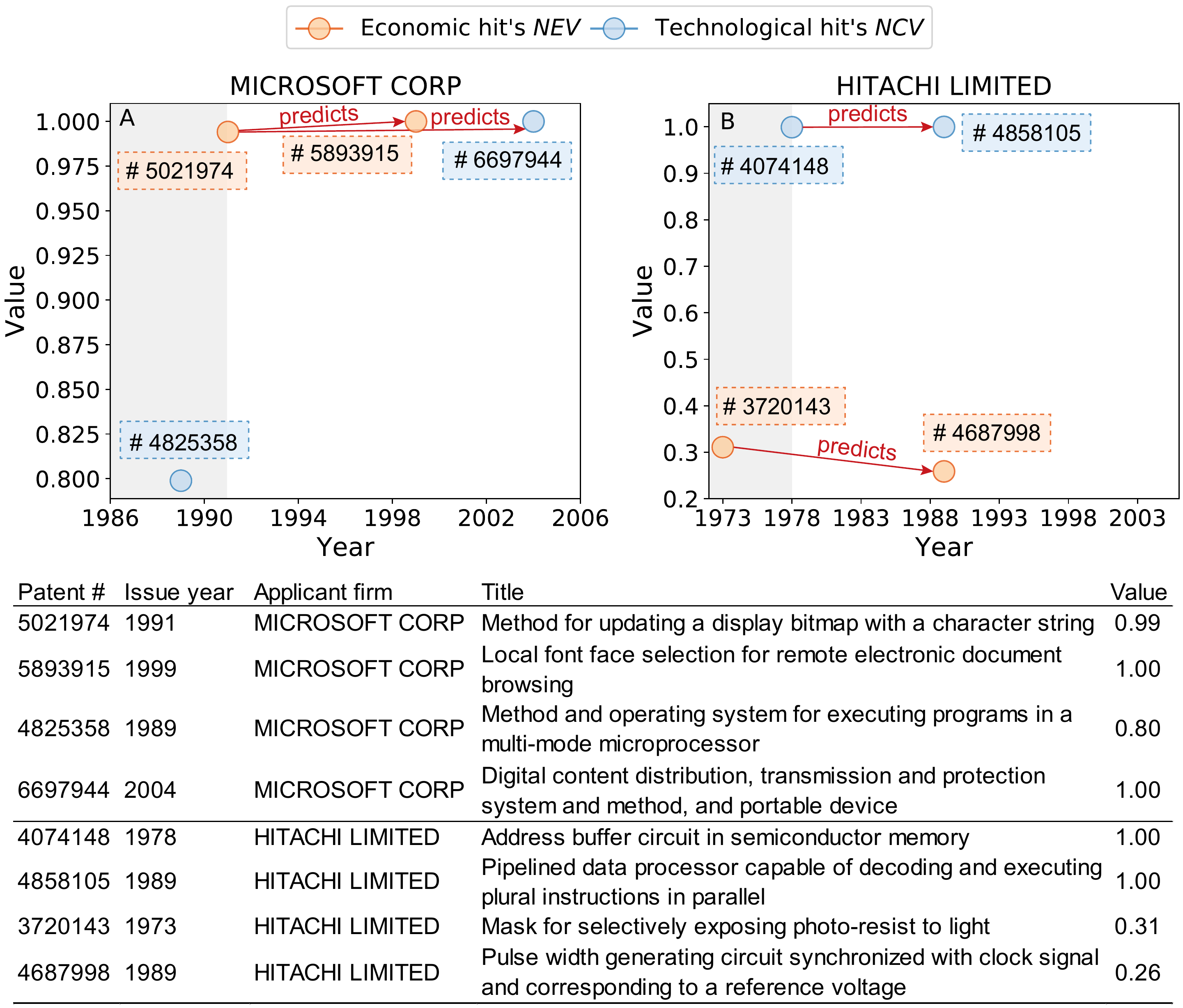}
\caption{\textbf{Example of two firms in terms of predicting.} 
(A) \textit{Microsoft} exhibits a high-value economic hit in both the early and late stage. During the early period (gray area), \textit{Microsoft}'s economic hit (panel A) was at a high level and its technological hit was at a low level; in the subsequent years (white area), the company produced both high-economic-value patents and high-technological value patents. (B) \textit{Hitachi Limited} exhibits high-value technological hits, but low-value economic hits, in both early and late stage. The table provides the details of these hit patents.}
\label{fig:Predict-final-example}
\end{figure}

We find that a firm's early economic value is the strongest predictor of both high-economic value firms and, more surprisingly, high-technological value firms in the future (see Fig.~\ref{fig:Predict-final-removeEarly} for results and Fig.~\ref{fig:Predict-final-example} for examples). 
By considering classifiers with $z\%=5\%$, the precision of the classifier based on early economic value reaches $51.6\%$ and $31.9\%$ for the prediction of high economic and technological value firms in the future (10.3-fold and 6.4-fold increase compared with a random classifier, respectively), as opposed to the smaller precision of the classifier based on early technological value ($2.8$-fold and $3.4$-fold increase compared to a random classifier, respectively:
see Figs.~\ref{fig:Predict-final-removeEarly}A and B; the results based on raw accuracy metrics are shown in Fig.~S6 in SI).
By summing over all possible values of $z\%$, the area under the precision-recall curve (\emph{AuPRC}) of the classifier based on early economic value is $4.12$ times and $1.72$ times larger than that of the classifier based on early technological value in the prediction of future high economic value firms and high technological firms, respectively (see Figs.~\ref{fig:Predict-final-removeEarly}C and D).  

The predictive power of firms' early economic value is substantially stronger than that of other predictors from the literature (such as early productivity and total citations), and significantly larger than that of a random classifier (see the dashed black lines in Fig.~\ref{fig:Predict-final-removeEarly}).
These conclusions are robust with respect to alternative choices of the prediction evaluation metric (see Figs. S6 in SI) and variations in the duration of the early window (see SI Fig.~S7) and subsequent window (see SI Fig.~S8). 
Combining all the early performance metrics via a binary logistic regression model can moderately improve the predictive accuracy only for the prediction of high-technological-value firms, at the cost of increasing model complexity (see Figs.~S6 in SI). For this reason, in the main text, we only show the result of single performance metrics.

Importantly, the stronger predictive power of early economic value holds as well when restricting the analysis to individual industrial sectors: by considering 10 macrosectors based on the first two digit of firms' Standard Industrial Classification (SIC) code\footnote{https://siccode.com/}, we find that the early economic value is the strongest predictor of future success for all $10$ industries except for the \emph{Transportation \& Public Utilities} sector (see SI, Fig.~S13 for details). This exception might occur because in this sector, the economic value of generated research is a weaker determinant of governments' and agencies' investment decisions.

\subsection{Explaining predictability: Competitive advantage or fitness?}

Two underlying mechanisms could explain the observed predictive power of the economic value of a firm's early patents, which we refer to as \textit{competitive advantage} and \textit{fitness} mechanism, respectively. According to the \textit{competitive advantage} mechanism, in the long-term, a firm might succeed because of the economic value she derives from her early patents. According to this interpretation, early economic success might allow firms to invest more in research, produce more patents in the future, and as a consequence, have more attempts to produce higher-value hits.
This mechanism would align with the Matthew Effect found in many other systems~\cite{perc2014the}.
On the other hand, according to a \textit{fitness} mechanism, the value of a firm's early hits might be a manifestation of the firm's ability to produce successful research, which could be interpreted as the firm's fitness. This mechanism would align with recent theories on the success dynamics of scientists that assume that a researcher's ability to produce high-impact papers is constant over time~\cite{sinatra2016quantifying,wang2021science}.

Recent works on success predictability~\cite{hofman2017prediction} and the science of science~\cite{li2019early,wang2021science} pointed out that disentangling the two mechanism in observational data is challenging. A definitive answer would indeed require a randomized controlled experiment~\cite{salganik2019bit,wang2021science}, which is impossible in our case.
Nevertheless, to move an initial step toward disentangling the two mechanisms, we implement a widely-used technique to approximate randomized experiments in quantitative social sciences: matched pair analysis~\cite{li2019early,salganik2019bit}.

\begin{figure}[t]
\centering
\includegraphics[scale=0.38]{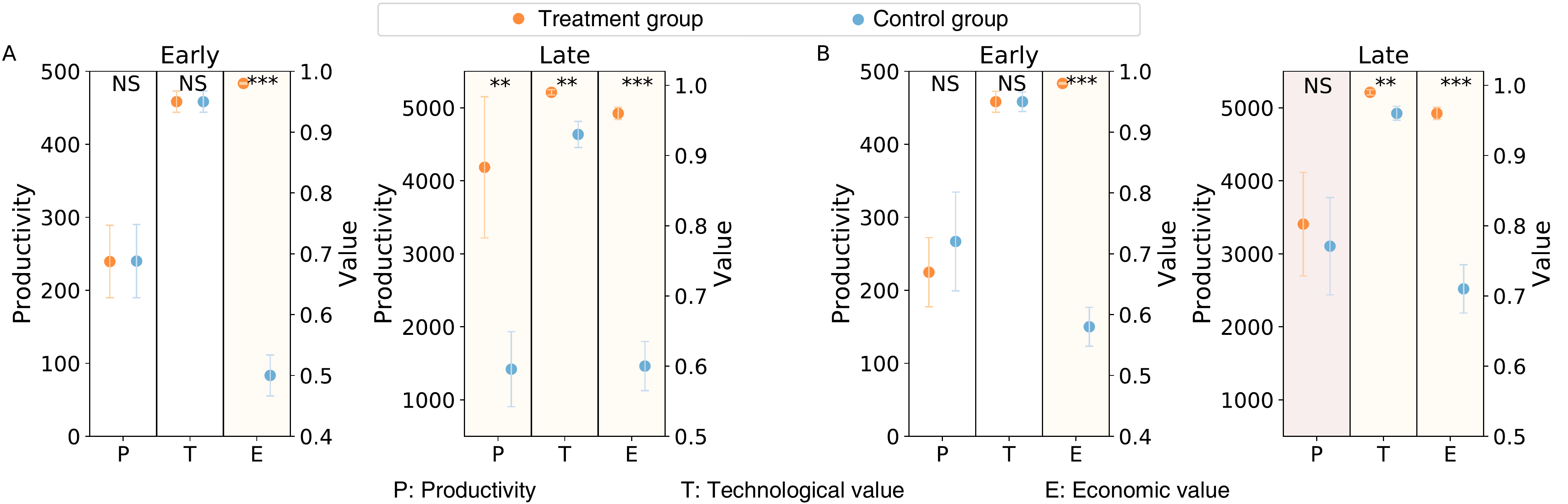}
\caption{\textbf{Quantifying the effect of early economic value: Matched pair analysis.} To quantify the impact of early economic value on subsequent success, we split the firms into a treatment and control group via propensity-score matching~\cite{rosenbaum1983the}. 
(A) Firms in the two groups have similar early productivity, technological value and the same SIC industry (the major 10 sectors), but significantly differ in early economic value. The results show firms in the ``treatment'' group with a larger early economic value exhibit a significant advantage in terms of future productivity, technological and economic value (see Table~S5 for the complete results). (B) We add late productivity to the covariates and find the difference in late technological and economic value is still significant
(see Table~S6 for the complete results), which provides evidence against the pure competitive advantage mechanism described in the main text.
Error bars denote standard errors of the mean. The p-value refers to the $t$-test. \emph{NS} represents not significant, $^{*}p<0.05$, $^{**}p<0.01$, $^{***}p<0.001$.
}
\label{fig:matched_simple}
\end{figure}

Specifically, we use propensity-score matching~\cite{rosenbaum1983the} (see Methods) to split pairs of same-industry firms (according to the 10 SIC macro-sectors) that are similar in terms of their early productivity and technological value, but differ significantly in terms of their early economic value (see Fig.~\ref{fig:matched_simple}A), which leads to two groups of firms. We find that firms in the ``treatment'' group with a larger early economic value exhibit a significant advantage in terms of number of issued patents ($+195\%, P<0.01$), technological value ($+6\%, P<0.01$) and economic value ($+60\%, P<0.001$) in the late window (see Fig.~\ref{fig:matched_simple}A and Table S5).
To verify whether a similar result would hold for early technological value, we perform a complementary experiment where the same-category firms' pairs to be split into two groups exhibit similar early productivity and economic value, but significantly different early technological value. In this scenario where only early technological value differentiates the two groups of firms, the difference in late technological value is significant ($+7\%, P<0.01$), but that in late economic value is not ($+10\%, P>0.05$, see SI, Table S5).  

These results might be naively interpreted as evidence in favor of the competitive advantage mechanism. But if firm's late success is entirely due to increased late productivity, success differences among the treated and controlled firms would disappear if the late number of issued patents is added to the covariates in the matching procedure.
However, we find that this is not the case: When adding the late productivity to the covariates, treated firms still exhibit a significant advantage over controlled firms in terms of technological value ($+3\%, P<0.01$) and economic value ($+35\%, P<0.001$) in the subsequent years (see Fig.~\ref{fig:matched_simple}B and Table S6 in SI). This finding rules out the possibility that firms with early high economic value succeed in the future merely because of increased late productivity. Taken together, these findings indicate that the competitive advantage mechanism alone cannot explain the observed predictive power, and the fitness mechanism may play a main role in the research success of firms.

\subsection{Early patent value predicts hidden gems}

\begin{figure}[t]
\centering
\includegraphics[scale=0.38]{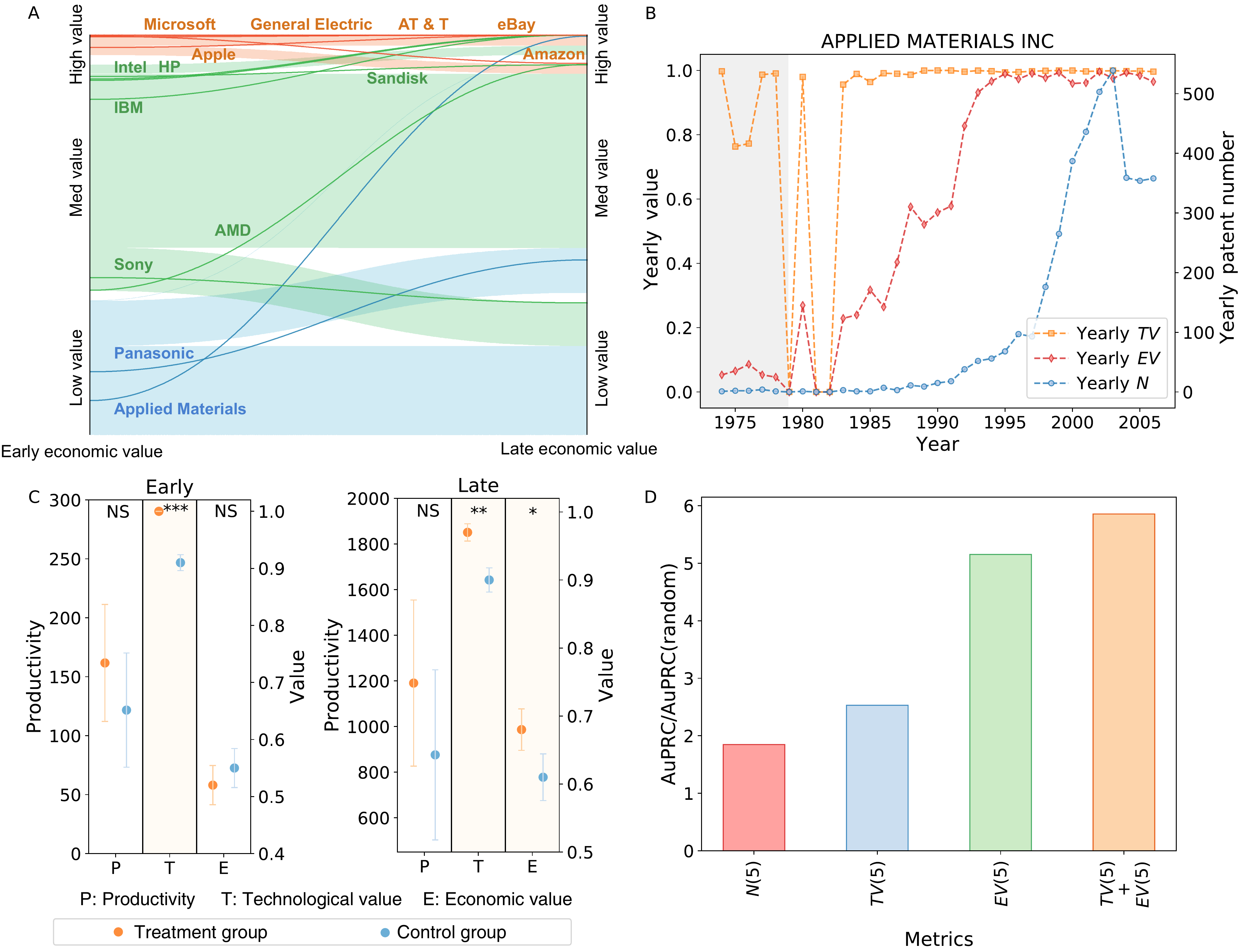}
\caption{\textbf{Predicting hidden gem firms.} (A) Illustration of the transition of firms' economic value level from the early stage (first 5 years) to the late stage (from the sixth year on). Among the 14 example firms in the figure, the \textit{predictable firms} include \textit{Microsoft}, \textit{General Electric}, \textit{AT\&T}, \textit{eBay} and \textit{Apple}; the \textit{hidden gems} are \textit{Intel}, \textit{HP}, \textit{IBM} and \textit{Applied Materials}. the \textit{declining firm} is \textit{Amazon}; the \textit{non-top firms} are \textit{SanDisk}, \textit{AMD}, \textit{Sony} and \textit{Panasonic}.
(B) An example of a hidden gem firm: \textit{Applied Materials Inc}. This panel shows the yearly economic (red line) and technological value (orange line) and yearly number of issued patents (blue line). Over the earliest 5 years (gray shadow), \textit{Applied Materials} was granted high technological-value patents, but stayed at low economic value. After the early period, the economic value of its patents steadily grew, until the firm's patents reached a high economic value. See the main text for a discussion and more examples in SI, Fig.~S9 .
(C) Matched pair analysis restricted to firms with medium or low economic value during the early phase. Firms in the treatment and control groups significantly differ in terms of their early technological value. Firms in the treatment group exhibit a significant advantage in terms of future economic and technological value (see Table~S7 for the complete results). Error bars stand for standard errors of the mean. The p-value refers to t-test. \emph{NS} represents not significant, $^{*}p<0.05$, $^{**}p<0.01$, $^{***}p<0.001$.
(D) Accuracy of simple classifiers for the early detection of hidden gem firms, in terms of their area under the precision-recall curve (\emph{AuPRC}) normalized by the \emph{AuPRC} for a random classifier. The best predictive performance is achieved by the combination of early economic and technological value.
}
\label{fig:transition}
\end{figure}

The observed predictability relies on the hypothesis that early top-firms are more likely to be among the high-value ones in the future. We refer to firms with early high economic value that maintain high economic value in the subsequent years as \textit{predictable firms}.
Despite the high precision of the resulting classifiers, there exist $2.3\%$ of firms that are not initially among the top-performing ones (i.e., top-$5\%$ by early economic value) and later end up among the top-$5\%$ (see Fig.~\ref{fig:transition}A). These ``low-to-high value'' firms, which we refer to as \textit{hidden gems}, are reminiscent of sleeping beauty papers in science~\cite{ke2015defining}: They are only able to be granted an economic hit after a relatively long time after their first patent issuance.
Here, we aim to quantify the early detectability of the set of hidden gem firms that transition from medium or low value to high value. 

\begin{table}[htbp]
\small
  \centering
  \caption{\textbf{A simple classification of firms.} }
    \begin{tabular}{|r|l|l|}
    \hline
     & Top late $EV$ & Non-top late $EV$ \\
    \hline
    Top early $EV$ (top $5\%$) & Predictable & Declining\\
    Non-top early $EV$ (bottom $95\%$) & Hidden gem & Non-top \\
    \hline
    \end{tabular}
  \label{tab:class}
\end{table}

Both \textit{predicable firms} and \textit{hidden gems} exhibit high late economic value. Besides, we refer to $92.7\%$ of firms that never reach high economic value as \textit{non-top firms}; to $2.3\%$ of firms that start from high economic value and descend to a lower value level as \textit{declining firms} (see Table~\ref{tab:class} for the classification of firms). 
We show the heterogeneous economic-value trajectories of 14 well-known firms in Fig.~\ref{fig:transition}A. Among them, \textit{Microsoft}, \textit{General Electric}, \textit{AT\&T}, \textit{eBay} and \textit{Apple} maintained a high value (predictable firms according to our definition), while \textit{Amazon} fell from high to medium value (declining firm). By contrast, \textit{Intel}, \textit{IBM}, and \textit{HP} went up from medium to high value, and \textit{Applied Materials} rose from low to high value; these four firms are hidden gems according to our definition.

\textit{Applied Materials} is an outstanding example of hidden gems. The firm was unable to produce high economic-value patents within its earliest $5$ years of patenting activity, although it was granted high-technological value patents in the early stage.
After 1982, its economic value exhibited a steady growth, and subsequently, the firm became able to produce high economic-value patents (see Fig.~\ref{fig:transition}B and Fig.~S9 in SI for more examples).
This transition is reflected in the company's history. \textit{Applied Materials} went public in 1972. In the subsequent few years, the company followed a diversified business strategy. During this period, its technological value was high, while its economic value was low. In 1976, it changed CEO and refocused to its core business of semiconductor manufacturing equipment\footnote{\url{https://en.wikipedia.org/wiki/Applied_Materials}}. After that, its economic value rapidly increased, whereas its technological value stayed at a high level. At the time of writing, the company is a global leader in its core industry. 

The existence of hidden gems raises the question: Are they predictable? The \textit{Applied Materials} example suggests that early high technological value might predict transitions from low or medium early economic value to high late economic value. 
We confirm this conjecture in two ways. First, we compare the early technological value for four groups of firms with distinct economic value dynamics (see SI, Fig.~S10). We find the average $TV(5)$ for hidden gem firms is 0.957 (s.e.m. 0.009), which is markedly larger than that for declining firms (0.939 (s.e.m. 0.022)), non-top firms (0.876 (s.e.m. 0.003)), and even slightly larger than predictable firms (0.954 (s.e.m. 0.016)). 
Subsequently, we perform a matched pair analysis in which we only consider firms with non-top early economic value, and the early technological value is used to split pairs of firms among a treatment and control group.
Among same-industry firms with similar non-top early economic value and early productivity, those with high early technological value exhibit $10\%$--higher late economic and technological value than those do not (see Fig.~\ref{fig:transition}C and SI, Table S7). 
These findings indicate that among firms with non-top early economic value, an early advantage in early technological value translates into a late advantage in terms of economic value.

We further evaluate our ability to early detect the hidden gems via their early economic and technological value.
To this end, we measure firms' economic and technological value within the earliest $5$ years, and we evaluate the predictive performance of a Na\"ive Bayes classifier that classifies a firm as a hidden gem if and only if it is among the top-$z\%$ by a given metric, where $z$ is a parameter that can be tuned to achieve a desired value of recall. We consider various performance metrics, including early productivity, $N(5)$, early economic value, $EV(5)$, early technological value, $TV(5)$, and the sum of early economic and technological value, $EV(5)+TV(5)$.

We find that the $EV(5)$ alone achieves a $5.1$ fold increase in \emph{AuPRC} compared to a random classifier (see Fig.~\ref{fig:transition}D). This signals that, unsurprisingly, firms that are nearer the top threshold in early stage are more likely to transition to high value. More interestingly, the $TV(5)$ alone achieves a $2.5$ fold increase in \emph{AuPRC} compared to a random classifier, and a combination of the $EV(5)$ and $TV(5)$ achieves the most accurate predictions, leading to a $5.8$ fold increase in \emph{AuPRC} compared to a random classifier (see Fig.~\ref{fig:transition}D and Fig.~S10 in SI for shortening the duration of early window), which confirms the key role of early technological value in the transition to high economic value.

\subsection{The timing of firms' hit patents is not random}

\begin{figure}[t]
\centering
\includegraphics[scale=0.37]{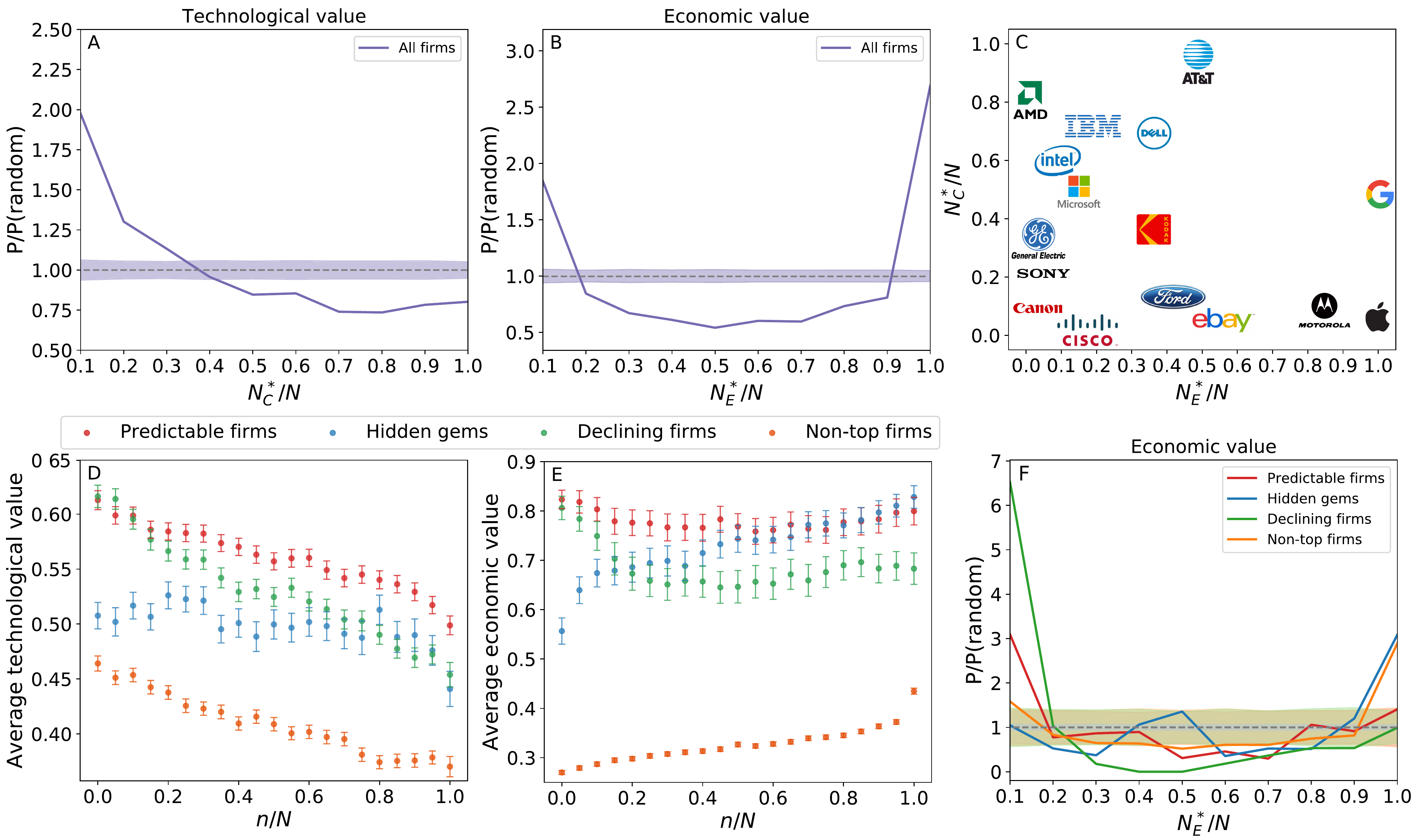}
\caption{\textbf{Heterogeneous patterns of research success over a firm's lifecycle.} (A) Probability distribution of $N_{C}^*/N$ for all analyzed firms, where $N_{C}^*/N\in[1/N,1]$ denotes the relative temporal position of the firms' technological hits (equal to $1/N$ or $1$ if the hit is the first or last issued patent of the firm, respectively). The overall decreasing trend significantly deviates from the expectation $P^{(random)}(N_{C}^*/N)=1$ for a randomized lifecycle. The shadow area shows the standard error of the results for 200 times randomized firms' lifecycles
(for each firm, patents' value scores are randomized, while the total number of issued patents is preserved).
(B) Probability distribution of $N_{E}^*/N$ for all analyzed firms, where $N_{E}^*/N\in[1/N,1]$ denotes the relative temporal position of the firms' economic hits. There a two-peaked distribution that significantly deviates from the expectation $P^{(random)}(N_{E}^*/N)=1$ for a randomized lifecycle.
(C) Illustration of $N_{C}^*$ \textit{vs}. $N_{E}^*$ for a sample of 16 famous firms, it shows firms like \textit{IBM}, \textit{AMD} and \textit{Intel} have an early economic hit but a relatively late technological hit, while for \textit{Apple} and \textit{Motorola}, the opposite is true, see details in SI, Table S4. (D) Average \emph{NCV} of firms' patents as a function of relative patent order for four groups of firms: predictable ($2.7\%$), hidden gems ($2.3\%$), non-top ($92.7\%$), and declining ($2.3\%$) firms. All groups of firms exhibit a declining trend. (E) Average \emph{NEV} of firms' patents as a function of relative patent order for the same four groups of firms. Whereas the average \emph{NEV} of patents by predictable and declining firms decline with patent order, the patents by 
non-top and hidden gem firms exhibit a clear increasing trend. (F) $P(N_{E}^*/N)$ for the same four groups of firms. The two peaks are manifestations of the heterogeneity of firms' value dynamics: predictable and declining firms only contribute to the early peak of the distribution, whereas hidden gem and non-top firms only contribute to the late peak.
}
\label{fig:economic}
\end{figure}

The observed predictability of firms' future hits from early patents motivates us to study the temporal dynamics of firms' patent value. Do firms tend to be granted their hits at the beginning of their research lifecycles? Or are firms' highest-value patents randomly distributed along a firm's lifecycle, similarly to the highest-impact works for scientists, artists, and musicians~\cite{sinatra2016quantifying,liu2018hot,janosov2020success}? How do these patterns differ for predictable and hidden gem firms?
We find that differently from results for individuals' creative works~\cite{sinatra2016quantifying,liu2018hot,janosov2020success}, the temporal position of a firm's hits is markedly non-random.

Specifically, we study the distributions $P(N_{C}^*/N)$ and $P(N_{E}^*/N)$ of the relative position of a firm's technological hit ($N^{*}_C$) and economic hit ($N^{*}_E$), respectively, compared to the firm's total number of issued patents, $N$~\cite{sinatra2016quantifying,janosov2020success}. Both types of hits are significantly more likely to occur among earliest patents than expected by chance, which is demonstrated by the left peaks of the two distributions (Figs.~\ref{fig:economic}A and B). 
The observed peaks cannot be explained by randomized patenting histories where for each firm, patents' value scores are randomized, while the total number of patents is preserved (see the shadowed area in Figs.~\ref{fig:economic}A and B)~\cite{sinatra2016quantifying}.
At the same time, whereas the probability to achieve the technological hit steadily decreases as a firm is granted more patents (Fig.~\ref{fig:economic}A), the probability to achieve the economic hit exhibits a second peak around the end of the lifecycle (at $N_E^*/N\sim 1$, see Fig.~\ref{fig:economic}B).
These results hold not only when considering all firms together, but also when considering separately high-value (top 5\% by their hits' value), medium-value (middle 60 \% by their hits' value), and low-value firms (bottom 35\% by their hits' value), and when considering firms from different industries, see Figs.~S11 and S12 in SI.

The heterogeneity of firms' hit position is well-illustrated by a few key case studies in Fig.~\ref{fig:economic}C (see Table S4 in SI for details). \emph{IBM} achieved its economic hit (about an integrated circuit with dielectric insulation) in 1976, whereas it achieved its technological hit significantly later (in 2002) with a patent on controlling access to shared storage devices. On the other hand, the \emph{Apple}'s technological hit appeared in 1992 (on a powered manager for a portable laptop computer), whereas its economic hit was issued substantially later (in 2006, on an improved method for generating multimedia non-linear effects).

The different behavior of $P(N_{C}^*/N)$ and $P(N_{E}^*/N)$ is a reflection of firms' heterogeneous value dynamics, which is linked to the predictive problem studied above. To demonstrate this point, we consider the previously-defined four groups of firms: predictable, hidden gem, non-top, and declining firms. For the four groups of firms, we find that the average technological value of their patents tends to steadily decrease over time (Fig.~\ref{fig:economic}D), which matches the higher probability of early appearance of technological hits. The only exception is the group of hidden gem firms, which exhibits a stabler trend. This suggests that the hidden gems' innovation ability does not diminish as they mature, which could be the key for their later transition.
By contrast, the dynamics of average economic value exhibits heterogeneous patterns. Whereas the average economic value of predictable and declining firms' patents tends to remain stable or decrease over the firms' lifecycles, the economic value of hidden gem and non-top firms sharply increases over time (Fig.~\ref{fig:economic}E). This different behavior is reflected in the behavior of $P(N_{E}^*/N)$: predictable and declining firms only contribute to the early peak, whereas hidden gem and non-top firms only contribute to the late peak (Fig.~\ref{fig:economic}F).

The observed early peak of $P(N_{C}^*/N)$ supports previous studies which claimed that newcomer firms are more likely to produce innovations of high technical quality~\cite{huergo2004does,balasubramanian2008firm}. This is because as firm age, they might gradually refine their innovation competence and organizational routines~\cite{sorensen2000aging}; in this phase, benefits from new technological advances might reduce~\cite{balasubramanian2008firm}. Hence, inventions by experienced firms are more likely to be the extension and improvement of their established innovative domains and technologies~\cite{sorensen2000aging}. Based on our previous results, we conjecture that the second economic peak of hidden gem firms might be due to the increasing ability of technologically-competitive firms to attract interest from the market. In some cases, like \textit{Applied Materials}, this might be due to organizational transformations. In other cases, the late peak might be due to factors that have been associated with late success in innovation research, including time-consuming knowledge acquisition~\cite{jones2009burden}, experience~\cite{sekara2018chaperone}, and reputation~\cite{petersen2014reputation}.

Taken together, these findings indicate that the firms' hits are not uniformly distributed along the firms' research lifecycles, which markedly differs from previous findings on the timing of success for scientists~\cite{sinatra2016quantifying}, artists~\cite{janosov2020success}, and musicians~\cite{janosov2020success}. This discrepancy indicates that previously-identified mechanisms (such as the $Q$-model~\cite{sinatra2016quantifying}) are unable to explain the value dynamics in firms' research lifecycles, which calls for new modeling approaches.

\section{Discussion}

Our work aims to uncover patterns behind the predictability and dynamics of firms' research success. By viewing each firm as a collection of its granted patents, we quantify firms' research success according to the economic and technological value of their patents in two periods (an early and a late stage). We demonstrate that the economic value of a firm's early patents is highly predictive of both the economic value and technological value of the firm's late patents. By contrast, surprisingly, the early technological value of a firm's patents is only predictive of the technological value of the firm's late patents.
Among firms with late patents of high economic value, we distinguish among ``predictable'' and ``hidden gem'' firms (namely, firms with and without high early economic value, respectively). We identify early signals that enable the early detection of the hidden gems. Specifically, for firms with relatively low economic value in the early stage, high early technological value can facilitate late economic value.
Besides, we find that predictable firms and hidden gem firms exhibit considerably different patterns of research success over time: The patents by predictable firms exhibit an approximately stable average economic value, whereas hidden gems' patents exhibit a sharply increasing average economic value.
Similarly, the economic hit patents by predictable firms tend to be among the earliest patents, whereas the opposite is true for hidden gems.
These results are strikingly different than those found for researchers in academia~\cite{sinatra2016quantifying,janosov2020success,wang2021science}, which indicates that models for the dynamics of human achievements are not applicable to firms' lifecycles.

The predictive power of the economic value of firms' early patents raises the question of whether early value determines future success (competitive advantage mechanism) or whether it unveils a firm's ``fitness'', i.e., its ability to produce high-value research.
A similar dilemma arose in recent studies on the predictability of scientists' future success from their early collaborations with already-established top-scientists~\cite{li2019early} and their early funding~\cite{bol2018the}, and on the predictability of online viral content from its early popularity momentum~\cite{shulman2016predictability}. The results of our matched pairs analysis provide evidence against the pure competitive advantage mechanism and in support of the role played by the fitness mechanism. The obtained findings suggest that firms' ability to be granted to successful research might be reflected in the economic value of their early patents. At the same time, a small set of hidden gems exhibit a slower progression toward research success, which could be early detected by analyzing both the technological and the economic value of their early patents.

To conclude, the obtained findings contribute to both the management literature on drivers of firms' performance~\cite{huergo2004does,gupta2013firm,guzman2015silicon}, and the recent cross-disciplinary literature on success in human activities~\cite{sinatra2016quantifying,liu2018hot,wu2019large,li2019early,wang2019early}.
While recent strides in the science of science have deepened our understanding of the success trajectories of academic researchers~\cite{sinatra2016quantifying,liu2018hot,wu2019large,li2019early,wang2019early,wang2021science}, our results provide the first step toward a quantitative understanding of the evolution of firms' research success from a complexity science standpoint. 
Beyond firms, the research approach developed here might find application to the prediction of the research success of other players, such as cities, regions, and nations. This can help forecast promising regions and companies, identify bottlenecks in research and innovation activities, and inform resource allocation strategies.

\section*{Acknowledgement}

This work is supported by the National Natural Science Foundation of China (Grant Nos. 61673150, 11622538). L.L. acknowledges the Science Strength Promotion Programme of UESTC, Chengdu. M.S.M. acknowledges financial support from the University of Zurich through the URPP Social Networks, the Swiss National Science Foundation (Grant No. 200021-182659), the UESTC professor research start-up (Grant No. ZYGX2018KYQD215). The findings, interpretations, and conclusions expressed in this paper are entirely those of the authors and do not necessarily reflect the views of the European Commission.

\section*{References}
\bibliography{bibliography}

\begin{thebibliography}{58}%
\makeatletter
\providecommand \@ifxundefined [1]{%
 \@ifx{#1\undefined}
}%
\providecommand \@ifnum [1]{%
 \ifnum #1\expandafter \@firstoftwo
 \else \expandafter \@secondoftwo
 \fi
}%
\providecommand \@ifx [1]{%
 \ifx #1\expandafter \@firstoftwo
 \else \expandafter \@secondoftwo
 \fi
}%
\providecommand \natexlab [1]{#1}%
\providecommand \enquote  [1]{``#1''}%
\providecommand \bibnamefont  [1]{#1}%
\providecommand \bibfnamefont [1]{#1}%
\providecommand \citenamefont [1]{#1}%
\providecommand \href@noop [0]{\@secondoftwo}%
\providecommand \href [0]{\begingroup \@sanitize@url \@href}%
\providecommand \@href[1]{\@@startlink{#1}\@@href}%
\providecommand \@@href[1]{\endgroup#1\@@endlink}%
\providecommand \@sanitize@url [0]{\catcode `\\12\catcode `\$12\catcode
  `\&12\catcode `\#12\catcode `\^12\catcode `\_12\catcode `\%12\relax}%
\providecommand \@@startlink[1]{}%
\providecommand \@@endlink[0]{}%
\providecommand \url  [0]{\begingroup\@sanitize@url \@url }%
\providecommand \@url [1]{\endgroup\@href {#1}{\urlprefix }}%
\providecommand \urlprefix  [0]{URL }%
\providecommand \Eprint [0]{\href }%
\providecommand \doibase [0]{http://dx.doi.org/}%
\providecommand \selectlanguage [0]{\@gobble}%
\providecommand \bibinfo  [0]{\@secondoftwo}%
\providecommand \bibfield  [0]{\@secondoftwo}%
\providecommand \translation [1]{[#1]}%
\providecommand \BibitemOpen [0]{}%
\providecommand \bibitemStop [0]{}%
\providecommand \bibitemNoStop [0]{.\EOS\space}%
\providecommand \EOS [0]{\spacefactor3000\relax}%
\providecommand \BibitemShut  [1]{\csname bibitem#1\endcsname}%
\let\auto@bib@innerbib\@empty
\bibitem [{idc(2019)}]{idc2019worldwide}%
  \BibitemOpen
  \bibfield  {title} {\enquote {\bibinfo {title} {Idc's worldwide artificial
  intelligence spending guide},}\ }\href@noop {} {\bibfield  {journal}
  {\bibinfo  {journal}
  {\url{https://www.idc.com/getdoc.jsp?containerId=IDC_P33198}}\ } (\bibinfo
  {year} {2019})}\BibitemShut {NoStop}%
\bibitem [{\citenamefont {Tunyasuvunakool}\ \emph {et~al.}(2021)\citenamefont
  {Tunyasuvunakool}, \citenamefont {Adler}, \citenamefont {Wu}, \citenamefont
  {Green}, \citenamefont {Zielinski}, \citenamefont {{\v{Z}}{\'\i}dek},
  \citenamefont {Bridgland}, \citenamefont {Cowie}, \citenamefont {Meyer},
  \citenamefont {Laydon} \emph {et~al.}}]{tunyasuvunakool2021highly}%
  \BibitemOpen
  \bibfield  {author} {\bibinfo {author} {\bibfnamefont {K.}~\bibnamefont
  {Tunyasuvunakool}}, \bibinfo {author} {\bibfnamefont {J.}~\bibnamefont
  {Adler}}, \bibinfo {author} {\bibfnamefont {Z.}~\bibnamefont {Wu}}, \bibinfo
  {author} {\bibfnamefont {T.}~\bibnamefont {Green}}, \bibinfo {author}
  {\bibfnamefont {M.}~\bibnamefont {Zielinski}}, \bibinfo {author}
  {\bibfnamefont {A.}~\bibnamefont {{\v{Z}}{\'\i}dek}}, \bibinfo {author}
  {\bibfnamefont {A.}~\bibnamefont {Bridgland}}, \bibinfo {author}
  {\bibfnamefont {A.}~\bibnamefont {Cowie}}, \bibinfo {author} {\bibfnamefont
  {C.}~\bibnamefont {Meyer}}, \bibinfo {author} {\bibfnamefont
  {A.}~\bibnamefont {Laydon}},  \emph {et~al.},\ }\bibfield  {title} {\enquote
  {\bibinfo {title} {Highly accurate protein structure prediction for the human
  proteome},}\ }\href@noop {} {\bibfield  {journal} {\bibinfo  {journal}
  {Nature}\ ,\ \bibinfo {pages} {1--9}} (\bibinfo {year} {2021})}\BibitemShut
  {NoStop}%
\bibitem [{\citenamefont {Le}\ \emph {et~al.}(2020)\citenamefont {Le},
  \citenamefont {Cramer}, \citenamefont {Chen},\ and\ \citenamefont
  {Mayhew}}]{le2020evolution}%
  \BibitemOpen
  \bibfield  {author} {\bibinfo {author} {\bibfnamefont {T.~T.}\ \bibnamefont
  {Le}}, \bibinfo {author} {\bibfnamefont {J.~P.}\ \bibnamefont {Cramer}},
  \bibinfo {author} {\bibfnamefont {R.}~\bibnamefont {Chen}}, \ and\ \bibinfo
  {author} {\bibfnamefont {S.}~\bibnamefont {Mayhew}},\ }\bibfield  {title}
  {\enquote {\bibinfo {title} {Evolution of the covid-19 vaccine development
  landscape},}\ }\href@noop {} {\bibfield  {journal} {\bibinfo  {journal}
  {Nature Reviews Drug Discovery}\ }\textbf {\bibinfo {volume} {19}},\ \bibinfo
  {pages} {667--8} (\bibinfo {year} {2020})}\BibitemShut {NoStop}%
\bibitem [{\citenamefont {Hauser}, \citenamefont {Tellis},\ and\ \citenamefont
  {Griffin}(2006)}]{hauser2006research}%
  \BibitemOpen
  \bibfield  {author} {\bibinfo {author} {\bibfnamefont {J.}~\bibnamefont
  {Hauser}}, \bibinfo {author} {\bibfnamefont {G.~J.}\ \bibnamefont {Tellis}},
  \ and\ \bibinfo {author} {\bibfnamefont {A.}~\bibnamefont {Griffin}},\
  }\bibfield  {title} {\enquote {\bibinfo {title} {Research on innovation: A
  review and agenda for marketing science},}\ }\href@noop {} {\bibfield
  {journal} {\bibinfo  {journal} {Marketing Science}\ }\textbf {\bibinfo
  {volume} {25}},\ \bibinfo {pages} {687--717} (\bibinfo {year}
  {2006})}\BibitemShut {NoStop}%
\bibitem [{\citenamefont {Nanda}, \citenamefont {Samila},\ and\ \citenamefont
  {Sorenson}(2020)}]{nanda2020persistent}%
  \BibitemOpen
  \bibfield  {author} {\bibinfo {author} {\bibfnamefont {R.}~\bibnamefont
  {Nanda}}, \bibinfo {author} {\bibfnamefont {S.}~\bibnamefont {Samila}}, \
  and\ \bibinfo {author} {\bibfnamefont {O.}~\bibnamefont {Sorenson}},\
  }\bibfield  {title} {\enquote {\bibinfo {title} {The persistent effect of
  initial success: Evidence from venture capital},}\ }\href@noop {} {\bibfield
  {journal} {\bibinfo  {journal} {Journal of Financial Economics}\ } (\bibinfo
  {year} {2020})}\BibitemShut {NoStop}%
\bibitem [{\citenamefont {Guzman}\ and\ \citenamefont
  {Stern}(2015)}]{guzman2015silicon}%
  \BibitemOpen
  \bibfield  {author} {\bibinfo {author} {\bibfnamefont {J.}~\bibnamefont
  {Guzman}}\ and\ \bibinfo {author} {\bibfnamefont {S.}~\bibnamefont {Stern}},\
  }\bibfield  {title} {\enquote {\bibinfo {title} {Where is silicon valley?}}\
  }\href@noop {} {\bibfield  {journal} {\bibinfo  {journal} {Science}\ }\textbf
  {\bibinfo {volume} {347}},\ \bibinfo {pages} {606--609} (\bibinfo {year}
  {2015})}\BibitemShut {NoStop}%
\bibitem [{\citenamefont {Sinatra}\ \emph {et~al.}(2016)\citenamefont
  {Sinatra}, \citenamefont {Wang}, \citenamefont {Deville}, \citenamefont
  {Song},\ and\ \citenamefont {Barab{\'a}si}}]{sinatra2016quantifying}%
  \BibitemOpen
  \bibfield  {author} {\bibinfo {author} {\bibfnamefont {R.}~\bibnamefont
  {Sinatra}}, \bibinfo {author} {\bibfnamefont {D.}~\bibnamefont {Wang}},
  \bibinfo {author} {\bibfnamefont {P.}~\bibnamefont {Deville}}, \bibinfo
  {author} {\bibfnamefont {C.}~\bibnamefont {Song}}, \ and\ \bibinfo {author}
  {\bibfnamefont {A.-L.}\ \bibnamefont {Barab{\'a}si}},\ }\bibfield  {title}
  {\enquote {\bibinfo {title} {Quantifying the evolution of individual
  scientific impact},}\ }\href@noop {} {\bibfield  {journal} {\bibinfo
  {journal} {Science}\ }\textbf {\bibinfo {volume} {354}},\ \bibinfo {pages}
  {aaf5239} (\bibinfo {year} {2016})}\BibitemShut {NoStop}%
\bibitem [{\citenamefont {Liu}\ \emph {et~al.}(2018)\citenamefont {Liu},
  \citenamefont {Wang}, \citenamefont {Sinatra}, \citenamefont {Giles},
  \citenamefont {Song},\ and\ \citenamefont {Wang}}]{liu2018hot}%
  \BibitemOpen
  \bibfield  {author} {\bibinfo {author} {\bibfnamefont {L.}~\bibnamefont
  {Liu}}, \bibinfo {author} {\bibfnamefont {Y.}~\bibnamefont {Wang}}, \bibinfo
  {author} {\bibfnamefont {R.}~\bibnamefont {Sinatra}}, \bibinfo {author}
  {\bibfnamefont {C.~L.}\ \bibnamefont {Giles}}, \bibinfo {author}
  {\bibfnamefont {C.}~\bibnamefont {Song}}, \ and\ \bibinfo {author}
  {\bibfnamefont {D.}~\bibnamefont {Wang}},\ }\bibfield  {title} {\enquote
  {\bibinfo {title} {Hot streaks in artistic, cultural, and scientific
  careers},}\ }\href@noop {} {\bibfield  {journal} {\bibinfo  {journal}
  {Nature}\ }\textbf {\bibinfo {volume} {559}},\ \bibinfo {pages} {396}
  (\bibinfo {year} {2018})}\BibitemShut {NoStop}%
\bibitem [{\citenamefont {Wang}, \citenamefont {Jones},\ and\ \citenamefont
  {Wang}(2019)}]{wang2019early}%
  \BibitemOpen
  \bibfield  {author} {\bibinfo {author} {\bibfnamefont {Y.}~\bibnamefont
  {Wang}}, \bibinfo {author} {\bibfnamefont {B.~F.}\ \bibnamefont {Jones}}, \
  and\ \bibinfo {author} {\bibfnamefont {D.}~\bibnamefont {Wang}},\ }\bibfield
  {title} {\enquote {\bibinfo {title} {Early-career setback and future career
  impact},}\ }\href@noop {} {\bibfield  {journal} {\bibinfo  {journal} {Nature
  Communications}\ }\textbf {\bibinfo {volume} {10}},\ \bibinfo {pages} {1--10}
  (\bibinfo {year} {2019})}\BibitemShut {NoStop}%
\bibitem [{\citenamefont {Yin}\ \emph {et~al.}(2019)\citenamefont {Yin},
  \citenamefont {Wang}, \citenamefont {Evans},\ and\ \citenamefont
  {Wang}}]{yin2019quantifying}%
  \BibitemOpen
  \bibfield  {author} {\bibinfo {author} {\bibfnamefont {Y.}~\bibnamefont
  {Yin}}, \bibinfo {author} {\bibfnamefont {Y.}~\bibnamefont {Wang}}, \bibinfo
  {author} {\bibfnamefont {J.~A.}\ \bibnamefont {Evans}}, \ and\ \bibinfo
  {author} {\bibfnamefont {D.}~\bibnamefont {Wang}},\ }\bibfield  {title}
  {\enquote {\bibinfo {title} {Quantifying the dynamics of failure across
  science, startups and security},}\ }\href@noop {} {\bibfield  {journal}
  {\bibinfo  {journal} {Nature}\ }\textbf {\bibinfo {volume} {575}},\ \bibinfo
  {pages} {190--194} (\bibinfo {year} {2019})}\BibitemShut {NoStop}%
\bibitem [{\citenamefont {Wang}\ and\ \citenamefont
  {Barab{\'a}si}(2021)}]{wang2021science}%
  \BibitemOpen
  \bibfield  {author} {\bibinfo {author} {\bibfnamefont {D.}~\bibnamefont
  {Wang}}\ and\ \bibinfo {author} {\bibfnamefont {A.-L.}\ \bibnamefont
  {Barab{\'a}si}},\ }\href@noop {} {\emph {\bibinfo {title} {Science of
  Science}}}\ (\bibinfo  {publisher} {In press},\ \bibinfo {year}
  {2021})\BibitemShut {NoStop}%
\bibitem [{\citenamefont {Fortunato}\ \emph {et~al.}(2018)\citenamefont
  {Fortunato}, \citenamefont {Bergstrom}, \citenamefont {B{\"o}rner},
  \citenamefont {Evans}, \citenamefont {Helbing}, \citenamefont
  {Milojevi{\'c}}, \citenamefont {Petersen}, \citenamefont {Radicchi},
  \citenamefont {Sinatra}, \citenamefont {Uzzi} \emph
  {et~al.}}]{fortunato2018science}%
  \BibitemOpen
  \bibfield  {author} {\bibinfo {author} {\bibfnamefont {S.}~\bibnamefont
  {Fortunato}}, \bibinfo {author} {\bibfnamefont {C.~T.}\ \bibnamefont
  {Bergstrom}}, \bibinfo {author} {\bibfnamefont {K.}~\bibnamefont
  {B{\"o}rner}}, \bibinfo {author} {\bibfnamefont {J.~A.}\ \bibnamefont
  {Evans}}, \bibinfo {author} {\bibfnamefont {D.}~\bibnamefont {Helbing}},
  \bibinfo {author} {\bibfnamefont {S.}~\bibnamefont {Milojevi{\'c}}}, \bibinfo
  {author} {\bibfnamefont {A.~M.}\ \bibnamefont {Petersen}}, \bibinfo {author}
  {\bibfnamefont {F.}~\bibnamefont {Radicchi}}, \bibinfo {author}
  {\bibfnamefont {R.}~\bibnamefont {Sinatra}}, \bibinfo {author} {\bibfnamefont
  {B.}~\bibnamefont {Uzzi}},  \emph {et~al.},\ }\bibfield  {title} {\enquote
  {\bibinfo {title} {Science of science},}\ }\href@noop {} {\bibfield
  {journal} {\bibinfo  {journal} {Science}\ }\textbf {\bibinfo {volume} {359}}
  (\bibinfo {year} {2018})}\BibitemShut {NoStop}%
\bibitem [{\citenamefont {Jaffe}\ and\ \citenamefont
  {De~Rassenfosse}(2019)}]{jaffe2019patent}%
  \BibitemOpen
  \bibfield  {author} {\bibinfo {author} {\bibfnamefont {A.~B.}\ \bibnamefont
  {Jaffe}}\ and\ \bibinfo {author} {\bibfnamefont {G.}~\bibnamefont
  {De~Rassenfosse}},\ }\bibfield  {title} {\enquote {\bibinfo {title} {Patent
  citation data in social science research: Overview and best practices},}\
  }in\ \href@noop {} {\emph {\bibinfo {booktitle} {Research Handbook on the
  Economics of Intellectual Property Law}}}\ (\bibinfo  {publisher} {Edward
  Elgar Publishing},\ \bibinfo {year} {2019})\BibitemShut {NoStop}%
\bibitem [{\citenamefont {Uzzi}\ \emph {et~al.}(2013)\citenamefont {Uzzi},
  \citenamefont {Mukherjee}, \citenamefont {Stringer},\ and\ \citenamefont
  {Jones}}]{uzzi2013atypical}%
  \BibitemOpen
  \bibfield  {author} {\bibinfo {author} {\bibfnamefont {B.}~\bibnamefont
  {Uzzi}}, \bibinfo {author} {\bibfnamefont {S.}~\bibnamefont {Mukherjee}},
  \bibinfo {author} {\bibfnamefont {M.}~\bibnamefont {Stringer}}, \ and\
  \bibinfo {author} {\bibfnamefont {B.}~\bibnamefont {Jones}},\ }\bibfield
  {title} {\enquote {\bibinfo {title} {Atypical combinations and scientific
  impact},}\ }\href@noop {} {\bibfield  {journal} {\bibinfo  {journal}
  {Science}\ }\textbf {\bibinfo {volume} {342}},\ \bibinfo {pages} {468--472}
  (\bibinfo {year} {2013})}\BibitemShut {NoStop}%
\bibitem [{\citenamefont {Mukherjee}\ \emph {et~al.}(2017)\citenamefont
  {Mukherjee}, \citenamefont {Romero}, \citenamefont {Jones},\ and\
  \citenamefont {Uzzi}}]{mukherjee2017nearly}%
  \BibitemOpen
  \bibfield  {author} {\bibinfo {author} {\bibfnamefont {S.}~\bibnamefont
  {Mukherjee}}, \bibinfo {author} {\bibfnamefont {D.~M.}\ \bibnamefont
  {Romero}}, \bibinfo {author} {\bibfnamefont {B.}~\bibnamefont {Jones}}, \
  and\ \bibinfo {author} {\bibfnamefont {B.}~\bibnamefont {Uzzi}},\ }\bibfield
  {title} {\enquote {\bibinfo {title} {The nearly universal link between the
  age of past knowledge and tomorrow’s breakthroughs in science and
  technology: The hotspot},}\ }\href@noop {} {\bibfield  {journal} {\bibinfo
  {journal} {Science Advances}\ }\textbf {\bibinfo {volume} {3}},\ \bibinfo
  {pages} {e1601315} (\bibinfo {year} {2017})}\BibitemShut {NoStop}%
\bibitem [{\citenamefont {Shi}\ and\ \citenamefont
  {Evans}(2019)}]{shi2019science}%
  \BibitemOpen
  \bibfield  {author} {\bibinfo {author} {\bibfnamefont {F.}~\bibnamefont
  {Shi}}\ and\ \bibinfo {author} {\bibfnamefont {J.}~\bibnamefont {Evans}},\
  }\bibfield  {title} {\enquote {\bibinfo {title} {Science and technology
  advance through surprise},}\ }\href@noop {} {\bibfield  {journal} {\bibinfo
  {journal} {arXiv preprint arXiv:1910.09370}\ } (\bibinfo {year}
  {2019})}\BibitemShut {NoStop}%
\bibitem [{\citenamefont {Pugliese}\ \emph
  {et~al.}(2019{\natexlab{a}})\citenamefont {Pugliese}, \citenamefont
  {Napolitano}, \citenamefont {Zaccaria},\ and\ \citenamefont
  {Pietronero}}]{pugliese2019coherent}%
  \BibitemOpen
  \bibfield  {author} {\bibinfo {author} {\bibfnamefont {E.}~\bibnamefont
  {Pugliese}}, \bibinfo {author} {\bibfnamefont {L.}~\bibnamefont
  {Napolitano}}, \bibinfo {author} {\bibfnamefont {A.}~\bibnamefont
  {Zaccaria}}, \ and\ \bibinfo {author} {\bibfnamefont {L.}~\bibnamefont
  {Pietronero}},\ }\bibfield  {title} {\enquote {\bibinfo {title} {Coherent
  diversification in corporate technological portfolios},}\ }\href@noop {}
  {\bibfield  {journal} {\bibinfo  {journal} {PLOS ONE}\ }\textbf {\bibinfo
  {volume} {14}} (\bibinfo {year} {2019}{\natexlab{a}})}\BibitemShut {NoStop}%
\bibitem [{\citenamefont {Pugliese}\ \emph
  {et~al.}(2019{\natexlab{b}})\citenamefont {Pugliese}, \citenamefont {Cimini},
  \citenamefont {Patelli}, \citenamefont {Zaccaria}, \citenamefont
  {Pietronero},\ and\ \citenamefont {Gabrielli}}]{pugliese2019unfolding}%
  \BibitemOpen
  \bibfield  {author} {\bibinfo {author} {\bibfnamefont {E.}~\bibnamefont
  {Pugliese}}, \bibinfo {author} {\bibfnamefont {G.}~\bibnamefont {Cimini}},
  \bibinfo {author} {\bibfnamefont {A.}~\bibnamefont {Patelli}}, \bibinfo
  {author} {\bibfnamefont {A.}~\bibnamefont {Zaccaria}}, \bibinfo {author}
  {\bibfnamefont {L.}~\bibnamefont {Pietronero}}, \ and\ \bibinfo {author}
  {\bibfnamefont {A.}~\bibnamefont {Gabrielli}},\ }\bibfield  {title} {\enquote
  {\bibinfo {title} {Unfolding the innovation system for the development of
  countries: co-evolution of science, technology and production},}\ }\href@noop
  {} {\bibfield  {journal} {\bibinfo  {journal} {Scientific Reports}\ }\textbf
  {\bibinfo {volume} {9}},\ \bibinfo {pages} {16440} (\bibinfo {year}
  {2019}{\natexlab{b}})}\BibitemShut {NoStop}%
\bibitem [{\citenamefont {Wang}, \citenamefont {Song},\ and\ \citenamefont
  {Barab{\'a}si}(2013)}]{wang2013quantifying}%
  \BibitemOpen
  \bibfield  {author} {\bibinfo {author} {\bibfnamefont {D.}~\bibnamefont
  {Wang}}, \bibinfo {author} {\bibfnamefont {C.}~\bibnamefont {Song}}, \ and\
  \bibinfo {author} {\bibfnamefont {A.-L.}\ \bibnamefont {Barab{\'a}si}},\
  }\bibfield  {title} {\enquote {\bibinfo {title} {Quantifying long-term
  scientific impact},}\ }\href@noop {} {\bibfield  {journal} {\bibinfo
  {journal} {Science}\ }\textbf {\bibinfo {volume} {342}},\ \bibinfo {pages}
  {127--132} (\bibinfo {year} {2013})}\BibitemShut {NoStop}%
\bibitem [{\citenamefont {Higham}\ \emph {et~al.}(2017)\citenamefont {Higham},
  \citenamefont {Governale}, \citenamefont {Jaffe},\ and\ \citenamefont
  {Z{\"u}licke}}]{higham2017fame}%
  \BibitemOpen
  \bibfield  {author} {\bibinfo {author} {\bibfnamefont {K.~W.}\ \bibnamefont
  {Higham}}, \bibinfo {author} {\bibfnamefont {M.}~\bibnamefont {Governale}},
  \bibinfo {author} {\bibfnamefont {A.}~\bibnamefont {Jaffe}}, \ and\ \bibinfo
  {author} {\bibfnamefont {U.}~\bibnamefont {Z{\"u}licke}},\ }\bibfield
  {title} {\enquote {\bibinfo {title} {Fame and obsolescence: Disentangling
  growth and aging dynamics of patent citations},}\ }\href@noop {} {\bibfield
  {journal} {\bibinfo  {journal} {Physical Review E}\ }\textbf {\bibinfo
  {volume} {95}},\ \bibinfo {pages} {042309} (\bibinfo {year}
  {2017})}\BibitemShut {NoStop}%
\bibitem [{\citenamefont {Wu}, \citenamefont {Wang},\ and\ \citenamefont
  {Evans}(2019)}]{wu2019large}%
  \BibitemOpen
  \bibfield  {author} {\bibinfo {author} {\bibfnamefont {L.}~\bibnamefont
  {Wu}}, \bibinfo {author} {\bibfnamefont {D.}~\bibnamefont {Wang}}, \ and\
  \bibinfo {author} {\bibfnamefont {J.~A.}\ \bibnamefont {Evans}},\ }\bibfield
  {title} {\enquote {\bibinfo {title} {Large teams develop and small teams
  disrupt science and technology},}\ }\href@noop {} {\bibfield  {journal}
  {\bibinfo  {journal} {Nature}\ }\textbf {\bibinfo {volume} {566}},\ \bibinfo
  {pages} {378--382} (\bibinfo {year} {2019})}\BibitemShut {NoStop}%
\bibitem [{\citenamefont {Kogan}\ \emph {et~al.}(2017)\citenamefont {Kogan},
  \citenamefont {Papanikolaou}, \citenamefont {Seru},\ and\ \citenamefont
  {Stoffman}}]{kogan2017technological}%
  \BibitemOpen
  \bibfield  {author} {\bibinfo {author} {\bibfnamefont {L.}~\bibnamefont
  {Kogan}}, \bibinfo {author} {\bibfnamefont {D.}~\bibnamefont {Papanikolaou}},
  \bibinfo {author} {\bibfnamefont {A.}~\bibnamefont {Seru}}, \ and\ \bibinfo
  {author} {\bibfnamefont {N.}~\bibnamefont {Stoffman}},\ }\bibfield  {title}
  {\enquote {\bibinfo {title} {{Technological Innovation, Resource Allocation,
  and Growth*}},}\ }\href {\doibase 10.1093/qje/qjw040} {\bibfield  {journal}
  {\bibinfo  {journal} {The Quarterly Journal of Economics}\ }\textbf {\bibinfo
  {volume} {132}},\ \bibinfo {pages} {665--712} (\bibinfo {year}
  {2017})}\BibitemShut {NoStop}%
\bibitem [{\citenamefont {Stoffman}, \citenamefont {Woeppel},\ and\
  \citenamefont {Yavuz}(2020)}]{stoffman2020small}%
  \BibitemOpen
  \bibfield  {author} {\bibinfo {author} {\bibfnamefont {N.}~\bibnamefont
  {Stoffman}}, \bibinfo {author} {\bibfnamefont {M.}~\bibnamefont {Woeppel}}, \
  and\ \bibinfo {author} {\bibfnamefont {M.~D.}\ \bibnamefont {Yavuz}},\
  }\bibfield  {title} {\enquote {\bibinfo {title} {Small innovators: No risk,
  no return},}\ }\href@noop {} {\bibfield  {journal} {\bibinfo  {journal}
  {Kelley School of Business Research Paper}\ } (\bibinfo {year}
  {2020})}\BibitemShut {NoStop}%
\bibitem [{Pat()}]{Patent_CRSP}%
  \BibitemOpen
  \href@noop {} {\enquote {\bibinfo {title} {Patent–crsp match, 1926-2017},}\
  }\bibinfo {howpublished} {Dropdox. Available at
  https://paper.dropbox.com/doc/Patent-CRSP-match-1926-2017-W3aHAj0Ce4CzKZayqCASj.
  Deposited 30 November 2019}\BibitemShut {NoStop}%
\bibitem [{\citenamefont {Hall}, \citenamefont {Jaffe},\ and\ \citenamefont
  {Trajtenberg}(2005)}]{hall2005market}%
  \BibitemOpen
  \bibfield  {author} {\bibinfo {author} {\bibfnamefont {B.~H.}\ \bibnamefont
  {Hall}}, \bibinfo {author} {\bibfnamefont {A.}~\bibnamefont {Jaffe}}, \ and\
  \bibinfo {author} {\bibfnamefont {M.}~\bibnamefont {Trajtenberg}},\
  }\bibfield  {title} {\enquote {\bibinfo {title} {Market value and patent
  citations},}\ }\href@noop {} {\bibfield  {journal} {\bibinfo  {journal} {RAND
  Journal of Economics}\ ,\ \bibinfo {pages} {16--38}} (\bibinfo {year}
  {2005})}\BibitemShut {NoStop}%
\bibitem [{\citenamefont {Waltman}(2016)}]{waltman2016review}%
  \BibitemOpen
  \bibfield  {author} {\bibinfo {author} {\bibfnamefont {L.}~\bibnamefont
  {Waltman}},\ }\bibfield  {title} {\enquote {\bibinfo {title} {A review of the
  literature on citation impact indicators},}\ }\href@noop {} {\bibfield
  {journal} {\bibinfo  {journal} {Journal of Informetrics}\ }\textbf {\bibinfo
  {volume} {10}},\ \bibinfo {pages} {365--391} (\bibinfo {year}
  {2016})}\BibitemShut {NoStop}%
\bibitem [{\citenamefont {Mariani}, \citenamefont {Medo},\ and\ \citenamefont
  {Lafond}(2019)}]{mariani2019early}%
  \BibitemOpen
  \bibfield  {author} {\bibinfo {author} {\bibfnamefont {M.~S.}\ \bibnamefont
  {Mariani}}, \bibinfo {author} {\bibfnamefont {M.}~\bibnamefont {Medo}}, \
  and\ \bibinfo {author} {\bibfnamefont {F.}~\bibnamefont {Lafond}},\
  }\bibfield  {title} {\enquote {\bibinfo {title} {Early identification of
  important patents: Design and validation of citation network metrics},}\
  }\href@noop {} {\bibfield  {journal} {\bibinfo  {journal} {Technological
  Forecasting and Social Change}\ }\textbf {\bibinfo {volume} {146}},\ \bibinfo
  {pages} {644--654} (\bibinfo {year} {2019})}\BibitemShut {NoStop}%
\bibitem [{\citenamefont {Li}\ \emph {et~al.}(2019)\citenamefont {Li},
  \citenamefont {Aste}, \citenamefont {Caccioli},\ and\ \citenamefont
  {Livan}}]{li2019early}%
  \BibitemOpen
  \bibfield  {author} {\bibinfo {author} {\bibfnamefont {W.}~\bibnamefont
  {Li}}, \bibinfo {author} {\bibfnamefont {T.}~\bibnamefont {Aste}}, \bibinfo
  {author} {\bibfnamefont {F.}~\bibnamefont {Caccioli}}, \ and\ \bibinfo
  {author} {\bibfnamefont {G.}~\bibnamefont {Livan}},\ }\bibfield  {title}
  {\enquote {\bibinfo {title} {Early coauthorship with top scientists predicts
  success in academic careers},}\ }\href@noop {} {\bibfield  {journal}
  {\bibinfo  {journal} {Nature Communications}\ }\textbf {\bibinfo {volume}
  {10}},\ \bibinfo {pages} {1--9} (\bibinfo {year} {2019})}\BibitemShut
  {NoStop}%
\bibitem [{\citenamefont {AlShebli}, \citenamefont {Rahwan},\ and\
  \citenamefont {Woon}(2018)}]{alshebli2018preeminence}%
  \BibitemOpen
  \bibfield  {author} {\bibinfo {author} {\bibfnamefont {B.~K.}\ \bibnamefont
  {AlShebli}}, \bibinfo {author} {\bibfnamefont {T.}~\bibnamefont {Rahwan}}, \
  and\ \bibinfo {author} {\bibfnamefont {W.~L.}\ \bibnamefont {Woon}},\
  }\bibfield  {title} {\enquote {\bibinfo {title} {The preeminence of ethnic
  diversity in scientific collaboration},}\ }\href@noop {} {\bibfield
  {journal} {\bibinfo  {journal} {Nature Communications}\ }\textbf {\bibinfo
  {volume} {9}},\ \bibinfo {pages} {1--10} (\bibinfo {year}
  {2018})}\BibitemShut {NoStop}%
\bibitem [{\citenamefont {Janosov}, \citenamefont {Battiston},\ and\
  \citenamefont {Sinatra}(2020)}]{janosov2020success}%
  \BibitemOpen
  \bibfield  {author} {\bibinfo {author} {\bibfnamefont {M.}~\bibnamefont
  {Janosov}}, \bibinfo {author} {\bibfnamefont {F.}~\bibnamefont {Battiston}},
  \ and\ \bibinfo {author} {\bibfnamefont {R.}~\bibnamefont {Sinatra}},\
  }\bibfield  {title} {\enquote {\bibinfo {title} {Success and luck in creative
  careers},}\ }\href {\doibase 10.1140/epjds/s13688-020-00227-w} {\bibfield
  {journal} {\bibinfo  {journal} {EPJ Data Science}\ }\textbf {\bibinfo
  {volume} {9}} (\bibinfo {year} {2020}),\
  10.1140/epjds/s13688-020-00227-w}\BibitemShut {NoStop}%
\bibitem [{\citenamefont {Strumsky}\ and\ \citenamefont
  {Lobo}(2015)}]{strumsky2015identifying}%
  \BibitemOpen
  \bibfield  {author} {\bibinfo {author} {\bibfnamefont {D.}~\bibnamefont
  {Strumsky}}\ and\ \bibinfo {author} {\bibfnamefont {J.}~\bibnamefont
  {Lobo}},\ }\bibfield  {title} {\enquote {\bibinfo {title} {Identifying the
  sources of technological novelty in the process of invention},}\ }\href@noop
  {} {\bibfield  {journal} {\bibinfo  {journal} {Research Policy}\ }\textbf
  {\bibinfo {volume} {44}},\ \bibinfo {pages} {1445--1461} (\bibinfo {year}
  {2015})}\BibitemShut {NoStop}%
\bibitem [{\citenamefont {Cremers}\ \emph {et~al.}(1999)\citenamefont
  {Cremers}, \citenamefont {Harhoff}, \citenamefont {Narin}, \citenamefont
  {Scherer},\ and\ \citenamefont {Vopel}}]{cremers1999citation}%
  \BibitemOpen
  \bibfield  {author} {\bibinfo {author} {\bibfnamefont {K.}~\bibnamefont
  {Cremers}}, \bibinfo {author} {\bibfnamefont {D.}~\bibnamefont {Harhoff}},
  \bibinfo {author} {\bibfnamefont {F.}~\bibnamefont {Narin}}, \bibinfo
  {author} {\bibfnamefont {F.}~\bibnamefont {Scherer}}, \ and\ \bibinfo
  {author} {\bibfnamefont {K.}~\bibnamefont {Vopel}},\ }\bibfield  {title}
  {\enquote {\bibinfo {title} {Citation frequency and the value of patented
  inventions},}\ }\href {\doibase 10.1162/003465399558265} {\bibfield
  {journal} {\bibinfo  {journal} {The Review of Economics and Statistics}\
  }\textbf {\bibinfo {volume} {81}},\ \bibinfo {pages} {511--515} (\bibinfo
  {year} {1999})}\BibitemShut {NoStop}%
\bibitem [{\citenamefont {Silverberg}\ and\ \citenamefont
  {Verspagen}(2007)}]{silverberg2007size}%
  \BibitemOpen
  \bibfield  {author} {\bibinfo {author} {\bibfnamefont {G.}~\bibnamefont
  {Silverberg}}\ and\ \bibinfo {author} {\bibfnamefont {B.}~\bibnamefont
  {Verspagen}},\ }\bibfield  {title} {\enquote {\bibinfo {title} {The size
  distribution of innovations revisited: an application of extreme value
  statistics to citation and value measures of patent significance},}\
  }\href@noop {} {\bibfield  {journal} {\bibinfo  {journal} {Journal of
  Econometrics}\ }\textbf {\bibinfo {volume} {139}},\ \bibinfo {pages}
  {318--339} (\bibinfo {year} {2007})}\BibitemShut {NoStop}%
\bibitem [{\citenamefont {Ahuja}\ and\ \citenamefont
  {Morris~Lampert}(2001)}]{ahuja2001entrepreneurship}%
  \BibitemOpen
  \bibfield  {author} {\bibinfo {author} {\bibfnamefont {G.}~\bibnamefont
  {Ahuja}}\ and\ \bibinfo {author} {\bibfnamefont {C.}~\bibnamefont
  {Morris~Lampert}},\ }\bibfield  {title} {\enquote {\bibinfo {title}
  {Entrepreneurship in the large corporation: A longitudinal study of how
  established firms create breakthrough inventions},}\ }\href@noop {}
  {\bibfield  {journal} {\bibinfo  {journal} {Strategic Management Journal}\
  }\textbf {\bibinfo {volume} {22}},\ \bibinfo {pages} {521--543} (\bibinfo
  {year} {2001})}\BibitemShut {NoStop}%
\bibitem [{\citenamefont {Fleming}\ and\ \citenamefont
  {Sorenson}(2003)}]{fleming2003navigating}%
  \BibitemOpen
  \bibfield  {author} {\bibinfo {author} {\bibfnamefont {L.}~\bibnamefont
  {Fleming}}\ and\ \bibinfo {author} {\bibfnamefont {O.}~\bibnamefont
  {Sorenson}},\ }\bibfield  {title} {\enquote {\bibinfo {title} {Navigating the
  technology landscape of innovation},}\ }\href@noop {} {\bibfield  {journal}
  {\bibinfo  {journal} {MIT Sloan Management Review}\ }\textbf {\bibinfo
  {volume} {44}},\ \bibinfo {pages} {15} (\bibinfo {year} {2003})}\BibitemShut
  {NoStop}%
\bibitem [{\citenamefont {Dunlap-Hinkler}, \citenamefont {Kotabe},\ and\
  \citenamefont {Mudambi}(2010)}]{dunlap2010story}%
  \BibitemOpen
  \bibfield  {author} {\bibinfo {author} {\bibfnamefont {D.}~\bibnamefont
  {Dunlap-Hinkler}}, \bibinfo {author} {\bibfnamefont {M.}~\bibnamefont
  {Kotabe}}, \ and\ \bibinfo {author} {\bibfnamefont {R.}~\bibnamefont
  {Mudambi}},\ }\bibfield  {title} {\enquote {\bibinfo {title} {A story of
  breakthrough versus incremental innovation: Corporate entrepreneurship in the
  global pharmaceutical industry},}\ }\href@noop {} {\bibfield  {journal}
  {\bibinfo  {journal} {Strategic Entrepreneurship Journal}\ }\textbf {\bibinfo
  {volume} {4}},\ \bibinfo {pages} {106--127} (\bibinfo {year}
  {2010})}\BibitemShut {NoStop}%
\bibitem [{\citenamefont {Srivastava}\ and\ \citenamefont
  {Gnyawali}(2011)}]{srivastava2011relational}%
  \BibitemOpen
  \bibfield  {author} {\bibinfo {author} {\bibfnamefont {M.~K.}\ \bibnamefont
  {Srivastava}}\ and\ \bibinfo {author} {\bibfnamefont {D.~R.}\ \bibnamefont
  {Gnyawali}},\ }\bibfield  {title} {\enquote {\bibinfo {title} {When do
  relational resources matter? leveraging portfolio technological resources for
  breakthrough innovation},}\ }\href@noop {} {\bibfield  {journal} {\bibinfo
  {journal} {Academy of Management Journal}\ }\textbf {\bibinfo {volume}
  {54}},\ \bibinfo {pages} {797--810} (\bibinfo {year} {2011})}\BibitemShut
  {NoStop}%
\bibitem [{\citenamefont {Ahuja}\ and\ \citenamefont
  {Katila}(2001)}]{ahuja2001technological}%
  \BibitemOpen
  \bibfield  {author} {\bibinfo {author} {\bibfnamefont {G.}~\bibnamefont
  {Ahuja}}\ and\ \bibinfo {author} {\bibfnamefont {R.}~\bibnamefont {Katila}},\
  }\bibfield  {title} {\enquote {\bibinfo {title} {Technological acquisitions
  and the innovation performance of acquiring firms: A longitudinal study},}\
  }\href@noop {} {\bibfield  {journal} {\bibinfo  {journal} {Strategic
  Management Journal}\ }\textbf {\bibinfo {volume} {22}},\ \bibinfo {pages}
  {197--220} (\bibinfo {year} {2001})}\BibitemShut {NoStop}%
\bibitem [{\citenamefont {Zhang}\ \emph {et~al.}(2020)\citenamefont {Zhang},
  \citenamefont {Zhang}, \citenamefont {Zhu},\ and\ \citenamefont
  {Liu}}]{zhang2020foot}%
  \BibitemOpen
  \bibfield  {author} {\bibinfo {author} {\bibfnamefont {S.}~\bibnamefont
  {Zhang}}, \bibinfo {author} {\bibfnamefont {N.}~\bibnamefont {Zhang}},
  \bibinfo {author} {\bibfnamefont {S.}~\bibnamefont {Zhu}}, \ and\ \bibinfo
  {author} {\bibfnamefont {F.}~\bibnamefont {Liu}},\ }\bibfield  {title}
  {\enquote {\bibinfo {title} {A foot in two camps or your undivided attention?
  the impact of intra-and inter-community collaboration on firm innovation
  performance},}\ }\href@noop {} {\bibfield  {journal} {\bibinfo  {journal}
  {Technology Analysis \& Strategic Management}\ }\textbf {\bibinfo {volume}
  {32}},\ \bibinfo {pages} {753--768} (\bibinfo {year} {2020})}\BibitemShut
  {NoStop}%
\bibitem [{\citenamefont {Trajtenberg}(1990)}]{trajtenberg1990penny}%
  \BibitemOpen
  \bibfield  {author} {\bibinfo {author} {\bibfnamefont {M.}~\bibnamefont
  {Trajtenberg}},\ }\bibfield  {title} {\enquote {\bibinfo {title} {A penny for
  your quotes: patent citations and the value of innovations},}\ }\href@noop {}
  {\bibfield  {journal} {\bibinfo  {journal} {The Rand Journal of Economics}\
  ,\ \bibinfo {pages} {172--187}} (\bibinfo {year} {1990})}\BibitemShut
  {NoStop}%
\bibitem [{\citenamefont {Turkina}, \citenamefont {Oreshkin},\ and\
  \citenamefont {Kali}(2019)}]{turkina2019regional}%
  \BibitemOpen
  \bibfield  {author} {\bibinfo {author} {\bibfnamefont {E.}~\bibnamefont
  {Turkina}}, \bibinfo {author} {\bibfnamefont {B.}~\bibnamefont {Oreshkin}}, \
  and\ \bibinfo {author} {\bibfnamefont {R.}~\bibnamefont {Kali}},\ }\bibfield
  {title} {\enquote {\bibinfo {title} {Regional innovation clusters and firm
  innovation performance: An interactionist approach},}\ }\href@noop {}
  {\bibfield  {journal} {\bibinfo  {journal} {Regional Studies}\ }\textbf
  {\bibinfo {volume} {53}},\ \bibinfo {pages} {1193--1206} (\bibinfo {year}
  {2019})}\BibitemShut {NoStop}%
\bibitem [{\citenamefont {Powers}(2011)}]{powers2011evaluation}%
  \BibitemOpen
  \bibfield  {author} {\bibinfo {author} {\bibfnamefont {D.}~\bibnamefont
  {Powers}},\ }\bibfield  {title} {\enquote {\bibinfo {title} {Evaluation: From
  predcision, recall and f-factor to roc, informedness, markedness \&
  correlation},}\ }\href@noop {} {\bibfield  {journal} {\bibinfo  {journal}
  {Journal of Machine Learning Technologies}\ }\textbf {\bibinfo {volume}
  {2}},\ \bibinfo {pages} {37--63} (\bibinfo {year} {2011})}\BibitemShut
  {NoStop}%
\bibitem [{\citenamefont {Perc}(2014)}]{perc2014the}%
  \BibitemOpen
  \bibfield  {author} {\bibinfo {author} {\bibfnamefont {M.}~\bibnamefont
  {Perc}},\ }\bibfield  {title} {\enquote {\bibinfo {title} {The matthew effect
  in empirical data.}}\ }\href@noop {} {\bibfield  {journal} {\bibinfo
  {journal} {Journal of the Royal Society Interface}\ }\textbf {\bibinfo
  {volume} {11}},\ \bibinfo {pages} {20140378--20140378} (\bibinfo {year}
  {2014})}\BibitemShut {NoStop}%
\bibitem [{\citenamefont {Hofman}, \citenamefont {Sharma},\ and\ \citenamefont
  {Watts}(2017)}]{hofman2017prediction}%
  \BibitemOpen
  \bibfield  {author} {\bibinfo {author} {\bibfnamefont {J.~M.}\ \bibnamefont
  {Hofman}}, \bibinfo {author} {\bibfnamefont {A.}~\bibnamefont {Sharma}}, \
  and\ \bibinfo {author} {\bibfnamefont {D.~J.}\ \bibnamefont {Watts}},\
  }\bibfield  {title} {\enquote {\bibinfo {title} {Prediction and explanation
  in social systems},}\ }\href@noop {} {\bibfield  {journal} {\bibinfo
  {journal} {Science}\ }\textbf {\bibinfo {volume} {355}},\ \bibinfo {pages}
  {486--488} (\bibinfo {year} {2017})}\BibitemShut {NoStop}%
\bibitem [{\citenamefont {Salganik}(2019)}]{salganik2019bit}%
  \BibitemOpen
  \bibfield  {author} {\bibinfo {author} {\bibfnamefont {M.~J.}\ \bibnamefont
  {Salganik}},\ }\href@noop {} {\emph {\bibinfo {title} {Bit by bit: Social
  research in the digital age}}}\ (\bibinfo  {publisher} {Princeton University
  Press},\ \bibinfo {year} {2019})\BibitemShut {NoStop}%
\bibitem [{\citenamefont {Rosenbaum}\ and\ \citenamefont
  {Rubin}(1983)}]{rosenbaum1983the}%
  \BibitemOpen
  \bibfield  {author} {\bibinfo {author} {\bibfnamefont {P.~R.}\ \bibnamefont
  {Rosenbaum}}\ and\ \bibinfo {author} {\bibfnamefont {D.~B.}\ \bibnamefont
  {Rubin}},\ }\bibfield  {title} {\enquote {\bibinfo {title} {The central role
  of the propensity score in observational studies for causal effects},}\
  }\href@noop {} {\bibfield  {journal} {\bibinfo  {journal} {Biometrika}\
  }\textbf {\bibinfo {volume} {70}},\ \bibinfo {pages} {41--55} (\bibinfo
  {year} {1983})}\BibitemShut {NoStop}%
\bibitem [{\citenamefont {Ke}\ \emph {et~al.}(2015)\citenamefont {Ke},
  \citenamefont {Ferrara}, \citenamefont {Radicchi},\ and\ \citenamefont
  {Flammini}}]{ke2015defining}%
  \BibitemOpen
  \bibfield  {author} {\bibinfo {author} {\bibfnamefont {Q.}~\bibnamefont
  {Ke}}, \bibinfo {author} {\bibfnamefont {E.}~\bibnamefont {Ferrara}},
  \bibinfo {author} {\bibfnamefont {F.}~\bibnamefont {Radicchi}}, \ and\
  \bibinfo {author} {\bibfnamefont {A.}~\bibnamefont {Flammini}},\ }\bibfield
  {title} {\enquote {\bibinfo {title} {Defining and identifying sleeping
  beauties in science},}\ }\href@noop {} {\bibfield  {journal} {\bibinfo
  {journal} {Proceedings of the National Academy of Sciences}\ }\textbf
  {\bibinfo {volume} {112}},\ \bibinfo {pages} {7426--7431} (\bibinfo {year}
  {2015})}\BibitemShut {NoStop}%
\bibitem [{\citenamefont {Huergo}\ and\ \citenamefont
  {Jaumandreu}(2004)}]{huergo2004does}%
  \BibitemOpen
  \bibfield  {author} {\bibinfo {author} {\bibfnamefont {E.}~\bibnamefont
  {Huergo}}\ and\ \bibinfo {author} {\bibfnamefont {J.}~\bibnamefont
  {Jaumandreu}},\ }\bibfield  {title} {\enquote {\bibinfo {title} {How does
  probability of innovation change with firm age?}}\ }\href@noop {} {\bibfield
  {journal} {\bibinfo  {journal} {Small Business Economics}\ }\textbf {\bibinfo
  {volume} {22}},\ \bibinfo {pages} {193--207} (\bibinfo {year}
  {2004})}\BibitemShut {NoStop}%
\bibitem [{\citenamefont {Balasubramanian}\ and\ \citenamefont
  {Lee}(2008)}]{balasubramanian2008firm}%
  \BibitemOpen
  \bibfield  {author} {\bibinfo {author} {\bibfnamefont {N.}~\bibnamefont
  {Balasubramanian}}\ and\ \bibinfo {author} {\bibfnamefont {J.}~\bibnamefont
  {Lee}},\ }\bibfield  {title} {\enquote {\bibinfo {title} {Firm age and
  innovation},}\ }\href@noop {} {\bibfield  {journal} {\bibinfo  {journal}
  {Industrial and Corporate Change}\ }\textbf {\bibinfo {volume} {17}},\
  \bibinfo {pages} {1019--1047} (\bibinfo {year} {2008})}\BibitemShut {NoStop}%
\bibitem [{\citenamefont {S{\o}rensen}\ and\ \citenamefont
  {Stuart}(2000)}]{sorensen2000aging}%
  \BibitemOpen
  \bibfield  {author} {\bibinfo {author} {\bibfnamefont {J.~B.}\ \bibnamefont
  {S{\o}rensen}}\ and\ \bibinfo {author} {\bibfnamefont {T.~E.}\ \bibnamefont
  {Stuart}},\ }\bibfield  {title} {\enquote {\bibinfo {title} {Aging,
  obsolescence, and organizational innovation},}\ }\href@noop {} {\bibfield
  {journal} {\bibinfo  {journal} {Administrative Science Quarterly}\ }\textbf
  {\bibinfo {volume} {45}},\ \bibinfo {pages} {81--112} (\bibinfo {year}
  {2000})}\BibitemShut {NoStop}%
\bibitem [{\citenamefont {Jones}(2009)}]{jones2009burden}%
  \BibitemOpen
  \bibfield  {author} {\bibinfo {author} {\bibfnamefont {B.~F.}\ \bibnamefont
  {Jones}},\ }\bibfield  {title} {\enquote {\bibinfo {title} {The burden of
  knowledge and the “death of the renaissance man”: Is innovation getting
  harder?}}\ }\href@noop {} {\bibfield  {journal} {\bibinfo  {journal} {The
  Review of Economic Studies}\ }\textbf {\bibinfo {volume} {76}},\ \bibinfo
  {pages} {283--317} (\bibinfo {year} {2009})}\BibitemShut {NoStop}%
\bibitem [{\citenamefont {Sekara}\ \emph {et~al.}(2018)\citenamefont {Sekara},
  \citenamefont {Deville}, \citenamefont {Ahnert}, \citenamefont
  {Barab{\'a}si}, \citenamefont {Sinatra},\ and\ \citenamefont
  {Lehmann}}]{sekara2018chaperone}%
  \BibitemOpen
  \bibfield  {author} {\bibinfo {author} {\bibfnamefont {V.}~\bibnamefont
  {Sekara}}, \bibinfo {author} {\bibfnamefont {P.}~\bibnamefont {Deville}},
  \bibinfo {author} {\bibfnamefont {S.~E.}\ \bibnamefont {Ahnert}}, \bibinfo
  {author} {\bibfnamefont {A.-L.}\ \bibnamefont {Barab{\'a}si}}, \bibinfo
  {author} {\bibfnamefont {R.}~\bibnamefont {Sinatra}}, \ and\ \bibinfo
  {author} {\bibfnamefont {S.}~\bibnamefont {Lehmann}},\ }\bibfield  {title}
  {\enquote {\bibinfo {title} {The chaperone effect in scientific
  publishing},}\ }\href@noop {} {\bibfield  {journal} {\bibinfo  {journal}
  {Proceedings of the National Academy of Sciences}\ }\textbf {\bibinfo
  {volume} {115}},\ \bibinfo {pages} {12603--12607} (\bibinfo {year}
  {2018})}\BibitemShut {NoStop}%
\bibitem [{\citenamefont {Petersen}\ \emph {et~al.}(2014)\citenamefont
  {Petersen}, \citenamefont {Fortunato}, \citenamefont {Pan}, \citenamefont
  {Kaski}, \citenamefont {Penner}, \citenamefont {Rungi}, \citenamefont
  {Riccaboni}, \citenamefont {Stanley},\ and\ \citenamefont
  {Pammolli}}]{petersen2014reputation}%
  \BibitemOpen
  \bibfield  {author} {\bibinfo {author} {\bibfnamefont {A.~M.}\ \bibnamefont
  {Petersen}}, \bibinfo {author} {\bibfnamefont {S.}~\bibnamefont {Fortunato}},
  \bibinfo {author} {\bibfnamefont {R.~K.}\ \bibnamefont {Pan}}, \bibinfo
  {author} {\bibfnamefont {K.}~\bibnamefont {Kaski}}, \bibinfo {author}
  {\bibfnamefont {O.}~\bibnamefont {Penner}}, \bibinfo {author} {\bibfnamefont
  {A.}~\bibnamefont {Rungi}}, \bibinfo {author} {\bibfnamefont
  {M.}~\bibnamefont {Riccaboni}}, \bibinfo {author} {\bibfnamefont {H.~E.}\
  \bibnamefont {Stanley}}, \ and\ \bibinfo {author} {\bibfnamefont
  {F.}~\bibnamefont {Pammolli}},\ }\bibfield  {title} {\enquote {\bibinfo
  {title} {Reputation and impact in academic careers},}\ }\href@noop {}
  {\bibfield  {journal} {\bibinfo  {journal} {Proceedings of the National
  Academy of Sciences}\ }\textbf {\bibinfo {volume} {111}},\ \bibinfo {pages}
  {15316--15321} (\bibinfo {year} {2014})}\BibitemShut {NoStop}%
\bibitem [{\citenamefont {Bol}, \citenamefont {De~Vaan},\ and\ \citenamefont
  {De~Rijt}(2018)}]{bol2018the}%
  \BibitemOpen
  \bibfield  {author} {\bibinfo {author} {\bibfnamefont {T.}~\bibnamefont
  {Bol}}, \bibinfo {author} {\bibfnamefont {M.}~\bibnamefont {De~Vaan}}, \ and\
  \bibinfo {author} {\bibfnamefont {A.~V.}\ \bibnamefont {De~Rijt}},\
  }\bibfield  {title} {\enquote {\bibinfo {title} {The matthew effect in
  science funding},}\ }\href@noop {} {\bibfield  {journal} {\bibinfo  {journal}
  {Proceedings of the National Academy of Sciences}\ }\textbf {\bibinfo
  {volume} {115}},\ \bibinfo {pages} {4887--4890} (\bibinfo {year}
  {2018})}\BibitemShut {NoStop}%
\bibitem [{\citenamefont {Shulman}, \citenamefont {Sharma},\ and\ \citenamefont
  {Cosley}(2016)}]{shulman2016predictability}%
  \BibitemOpen
  \bibfield  {author} {\bibinfo {author} {\bibfnamefont {B.}~\bibnamefont
  {Shulman}}, \bibinfo {author} {\bibfnamefont {A.}~\bibnamefont {Sharma}}, \
  and\ \bibinfo {author} {\bibfnamefont {D.}~\bibnamefont {Cosley}},\
  }\bibfield  {title} {\enquote {\bibinfo {title} {Predictability of
  popularity: Gaps between prediction and understanding},}\ }in\ \href@noop {}
  {\emph {\bibinfo {booktitle} {Tenth International AAAI Conference on Web and
  Social Media}}}\ (\bibinfo {year} {2016})\BibitemShut {NoStop}%
\bibitem [{\citenamefont {Gupta}, \citenamefont {Guha},\ and\ \citenamefont
  {Krishnaswami}(2013)}]{gupta2013firm}%
  \BibitemOpen
  \bibfield  {author} {\bibinfo {author} {\bibfnamefont {P.~D.}\ \bibnamefont
  {Gupta}}, \bibinfo {author} {\bibfnamefont {S.}~\bibnamefont {Guha}}, \ and\
  \bibinfo {author} {\bibfnamefont {S.~S.}\ \bibnamefont {Krishnaswami}},\
  }\bibfield  {title} {\enquote {\bibinfo {title} {Firm growth and its
  determinants},}\ }\href@noop {} {\bibfield  {journal} {\bibinfo  {journal}
  {Journal of Innovation and Entrepreneurship}\ }\textbf {\bibinfo {volume}
  {2}},\ \bibinfo {pages} {15} (\bibinfo {year} {2013})}\BibitemShut {NoStop}%
\bibitem [{\citenamefont {Leydesdorff}\ \emph {et~al.}(2011)\citenamefont
  {Leydesdorff}, \citenamefont {Bornmann}, \citenamefont {Mutz},\ and\
  \citenamefont {Opthof}}]{leydesdorff2011turning}%
  \BibitemOpen
  \bibfield  {author} {\bibinfo {author} {\bibfnamefont {L.}~\bibnamefont
  {Leydesdorff}}, \bibinfo {author} {\bibfnamefont {L.}~\bibnamefont
  {Bornmann}}, \bibinfo {author} {\bibfnamefont {R.}~\bibnamefont {Mutz}}, \
  and\ \bibinfo {author} {\bibfnamefont {T.}~\bibnamefont {Opthof}},\
  }\bibfield  {title} {\enquote {\bibinfo {title} {Turning the tables on
  citation analysis one more time: Principles for comparing sets of
  documents},}\ }\href@noop {} {\bibfield  {journal} {\bibinfo  {journal}
  {Journal of the American Society for Information Science and technology}\
  }\textbf {\bibinfo {volume} {62}},\ \bibinfo {pages} {1370--1381} (\bibinfo
  {year} {2011})}\BibitemShut {NoStop}%
\bibitem [{\citenamefont {Carley}, \citenamefont {Hedge},\ and\ \citenamefont
  {Marco}(2015)}]{carley2015probability}%
  \BibitemOpen
  \bibfield  {author} {\bibinfo {author} {\bibfnamefont {M.}~\bibnamefont
  {Carley}}, \bibinfo {author} {\bibfnamefont {D.}~\bibnamefont {Hedge}}, \
  and\ \bibinfo {author} {\bibfnamefont {A.}~\bibnamefont {Marco}},\ }\bibfield
   {title} {\enquote {\bibinfo {title} {What is the probability of receiving a
  us patent},}\ }\href@noop {} {\bibfield  {journal} {\bibinfo  {journal} {Yale
  JL \& Tech.}\ }\textbf {\bibinfo {volume} {17}},\ \bibinfo {pages} {203}
  (\bibinfo {year} {2015})}\BibitemShut {NoStop}%
\end{thebibliography}%

\clearpage

\section{Methods}

\subsection{The USPTO dataset \label{sec:data}}

We analyze the $2,458,402$ patents granted to $7,440$ firms by the United States Patent and Trademark Office from $1926-2017$~\cite{Patent_CRSP}. The average number of patents per firm is $330.6$, and the largest number is $123,220$ (granted to \textit{IBM}). 
For each patent, the dataset includes an ID, date of filling and grant date (we employ the latter date), IDs of the applicant firms, and the list of its cited patents. 
Note that this dataset only includes patents whose assignee has been matched to a firm in CRSP (Center for Research in Security Prices), so that each patent's applicant is a firm listed in the US stock market. All other patents are not included in the original data.

A potential issue when measuring the number of citations received by a patent is that it might be unreliable for patents issued near the end of the data. To prevent this issue, we make a conservative choice and limit our analysis to patents issued up to $2006$ (and their applicant firms). In this way, patents' citation counts are measured over a time-window of at least 10 years, thereby avoiding short-term fluctuations (results stay qualitatively the same if we consider patents issued up to $2011$). 
At the same time, we are interested in firms with a sufficiently productive research activity. For this reason, in the main text, we limit the firm-level analysis to firms that have at least $15$ patents, which includes $2,819$ firms. In the SI, we show that our main results are qualitatively the same when filtering the firms based on their number of years of activity, see Fig.~S16 in SI.

\subsection{Measuring technological and economic value}

\subsubsection{Patents' technological value \label{sec:method-YCRI}}

Citation count is traditionally used to gauge the scientific impact of papers~\cite{waltman2016review} and patents~\cite{jaffe2019patent}.
However, citation count should be used with caution because of its biases that make it unreliable to compare patents issued in different years~\cite{mariani2019early}. 
To fairly compare the impact of patents issued at different times, inspired by the percentile ranks in \cite{leydesdorff2011turning}, we measure patent $i$'s normalized citation value (\emph{NCV}) as $i$'s relative ranking position by citation count compared to all the patent issued in the same year as $i$. The definition reads 
\begin{equation}
    NCV_i=1-r_i/N(t_i),
\end{equation}
where $N(t_i)$ is the number of patents issued in the same year $t_i$ as patent $i$, and $r_i$ denotes the ranking of $i$ by citation count among the $N(t_i)$ patents of the same age ($r_i=1$ if $i$ is the top patent; $r_i=N(t_i)$ if $i$ is the last one, which correspond sto $NCV_i=1-N(t_i)^{-1}$ and $NCV_i=0$, respectively. Note that all tied values will be assigned the average of the rankings). Therefore, the resulting score $NCV_i\in [0,1)$ is close to one (zero) for high-value (low-value) patents. Crucially, differently from the rankings by citation count (see Fig.~\ref{fig:timebias}) and $C_{10}$ (i.e., citation count restricted to the first 10 years after the patent issuance, see SI, Fig.~S1)~\cite{sinatra2016quantifying}, the ranking by \emph{NCV} is consistent with an age-unbiased ranking (see Fig.~\ref{fig:timebias}).

\subsubsection{Patents' economic value\label{sec:method-YERI}}

In view of patent issuance conveys important information to the market, previous studies~\cite{stoffman2020small,kogan2017technological} estimated the US patents' economic value $\xi$ based on the movements of the applicant firm's stock prices over the days after the patents were issued. 
To observe the market's reaction to the patent grant, the authors adopted a two-day time window after the patent issuance based on the finding that firm's share turnover increases at most in the first two days after the patent issuance announcement.

To disentangle the component of firm return related to the patent's economic value from unrelated factors, they assumed that the idiosyncratic stock return $R$ for a given firm around the time window that its patent $j$ issued is,
\begin{equation}
    R_j=v_j+\epsilon_j,
\end{equation}
where $R$ equals to the firm's return minus the return on the market portfolio (to remove market movements), $v_j$ is the value of patent $j$, as a fraction of the firm's market capitalization and $\epsilon_j$ denotes the component of the firm's stock return that is unrelated to the patent.
Then, the economic value $\xi$ of patent $j$ is estimated as the product of the
estimate of the stock return due to the patent value times the market capitalization ($M_j$) of the applicant firm on the day prior to the patent issuance announcement:
\begin{equation}
    \xi_j=(1-\overline\pi)^{-1}\frac{1}{N_j}E\left[v_j|R_j\right]M_j,
\end{equation}
where $\overline\pi$ denotes the unconditional probability of a successful patent application, which is approximately 56\% according to patents filed between 1996 and 2005 and examined before
mid-2013 \cite{carley2015probability}. If on the same day of $j$ issued, $N_j$ patents are issued to the same firm, patent $j$ is assigned $1/N_j$ of the total value.
See \cite{stoffman2020small} for complete details of the estimation procedure. We use the ready-made estimated results provided by the authors which is available at \url{https://paper.dropbox.com/doc/Patent-CRSP-match-1926-2017-W3aHAj0Ce4CzKZayqCASj}.

The computation of the \emph{NEV} is similar to the \emph{NCV}'s one. We compare the $\xi$ value of patents issued in the same year, and patent $i$ will obtain a score $NEV_i=1-r_i/N(t_i)$, where $r_i/N(t_i)$ denotes $i$'s relative ranking position by $\xi$ among patents issued in the same year as $i$. Likewise, \emph{NEV} ranges in $[0,1)$.

\subsection{Matched pair analysis}
Matched pair analysis is a form of analysis in which each of the subjects in a treatment group is paired with each of those in a control group on the basis of matching covariates. 
This technique is widely used in medical and social research to evaluate the effect of a treatment, with the ease of implementation and comprehension.
We obtain such pairs via Propensity Score Matching~\cite{rosenbaum1983the}, where the propensity score is defined as the probability of treatment assignment conditional on baseline covariates.
We implement the matching by a Python package available at \url{http://www.kellieottoboni.com/pscore_match/} (with minor changes to support one-to-one matching). 

Take Fig.~\ref{fig:matched_simple}A as an example to explain the matching process. Firstly, we calculate propensity scores for each analyzed firm by applying a logistic regression where the covariates are early productivity and technological value and the dependent variable is whether the firm has top-5\% economic value in the early stage (1 if yes, 0 otherwise). Then each firm with top-5\% early economic value will be tried to match with one firm with non-top early economic value according to their propensity score, at the same time, we require the matching pairs to have identical SIC industry (the major 10 sectors).
Note that we use one-to-one match so that each firm with top early economic value will be matched with at most one firm with non-top early economic value. 
If the match succeeds (i.e. the two firms have close enough propensity score), the firm with top-5\% early economic value will be assigned to the treatment group, the other firm will be assigned to the control group. 
By going through all firms that have top-5\% economic value in the early time, we construct the treatment group and control group, and compare the subsequent performance of firms in the two groups.  

\section*{Supplementary Information}
Supplementary Information can be requested from S.X. (\url{xushuqi@std.uestc.edu.cn}).

\end{document}